\documentclass[aps,pre,twocolumn]{revtex4}
\usepackage{amsmath,bm,epsfig}
\usepackage{color}
\usepackage[normalem]{ulem}
\def\Fbox#1{\vskip1ex\hbox to 8.5cm{\hfil\fboxsep0.3cm\fbox{%
  \parbox{8.0cm}{#1}}\hfil}\vskip1ex\noindent}  

\newcommand{\B}[1]{{\bm{#1}}}
\newcommand{\C}[1]{{\mathcal{#1}}}    
\let \= \equiv \let\*\cdot \let\~\widetilde \let\-\overline

\begin{document}
\title{The Yield-Strain in Shear Banding Amorphous Solids}
\author{Ratul Dasgupta, H. George E. Hentschel$^*$ and  Itamar Procaccia}
\affiliation{Department of Chemical Physics, The Weizmann
 Institute of Science, Rehovot 76100, Israel.\\$^*$ Dept of Physics, Emory University, Atlanta GA. 30322}
\date{\today}
\begin{abstract}
In recent research it was found that the fundamental shear-localizing instability of amorphous solids under external strain, which eventually results in a shear band and failure, consists of a highly correlated array of Eshelby quadrupoles all having the same orientation and some density $\rho$. In this paper we calculate analytically the energy $E(\rho,\gamma)$ associated with such highly correlated structures as a function of the density $\rho$ and the external strain $\gamma$. We show that
for strains smaller than a characteristic strain $\gamma_Y$ the total
strain energy initially increases as the quadrupole density increases,
but that for strains larger than $\gamma_Y$ the energy monotonically
decreases with quadrupole density. We identify $\gamma_Y$ as the yield
strain. Its value, derived from values of the qudrupole strength based
on the atomistic model, agrees with that from the computed
stress-strain curves and broadly with experimental results.

\end{abstract}

\maketitle

\section{Introduction}
Amorphous solids are obtained when a glass-former is cooled below the glass transition \cite{08Che,06Dyr,09Cav}
to a state which on the one hand is amorphous, exhibiting liquid like organization of the constituents (atoms, molecules or polymers),
and on the other hand is a solid, reacting elastically (reversibly) to small strains. There is a large variety of experimental examples of such glassy systems, and theoretically there are many well studied models \cite{07BK,ML,09LP} based on point particles with a variety of inter-particle potentials that exhibit stable supercooled liquids phases which then solidify to an amorphous solid when cooled below the glass transition. Typically all these materials, both in the lab and on the computer, exhibit a so-called yield-stress above which the material fails to a plastic flow. In previous research  \cite{82SSH,02HFOLB,07SKLF,06TLB} it was pointed out that depending on the protocol of cooling to the glass state,
the plastic response of the system can be either via homogeneous flow or via shear bands. The former obtains typically when the quenching to the glass state is ``fast", whereas the latter when the quench is ``slow". In the latter case when the stress exceeds some  yield stress, the sample, rather than flowing homogeneously in a plastic flow, localizes all the shear in a plane that is at 45 degrees to the compressive stress axis, and then breaks along this plane \cite{08Che}.

In recent work \cite{12DHP} it was argued that the fundamental instability that gives rise to shear bands is the
appearance of highly correlated lines of Eshelby quadrupoles (and see below for precise definition) which organize the non-affine displacement field of an amorphous solid such that the shear is highly localized along a narrow band. How this fundamental instability results in macroscopic shear bands, why these appear in 45 degrees to the principal stress axis, and what determines the difference in plastic response between fastly and slowly quenched glasses are all subjects of this paper. We will also present an ab-initio calculation of the yield stress at which an amorphous solid is expected to response plastically with
shear localization.

In Sect. \ref{plastic} we review briefly the type of numerical simulations that we do and explain the basic facts about plasticity of amorphous solids. Section \ref{inclusion} exhibits the
fundamental solution of an Eshelby quadrupolar plastic event. This section is not particularly
new but is important for our purposes in setting the stage and the notation for the next Sect.
\ref{energy} in which we compute the energy of $\C N$ such Eshelby quadrupoles in the elastic medium. We show explicitly that as a function of the external strain (or the resulting stress)
there is a threshold value at which a bifurcation occurs. Below this value only isolated
Eshelby quadrupoles can appear in the system, leading to localized plastic events. Above this
threshold a density of such quadrupoles can appear, and when they do appear they are highly
correlated, preferring to organize on a line at 45 degrees to the principal shear direction.
In Sect. \ref{estimate} we present the analytic estimate of the yield strain, and demonstrate
a satisfactory agreement with the numerical simulations. Finally, in Sect. \ref{summary} we provide a summary of the most important results of the paper and offer a discussion of the road ahead.

\section{Plasticity in Amorphous Solids and Simulations}
\label{plastic}
As a background to the calculations in this paper we need to briefly review recent progress in understanding plasticity in amorphous solids \cite{ML,09LP,10HKLP,10KLP,12DKP}. Below we deal with 2-dimensional systems composed of $N$ point particles in an area $A$, characterized by a total energy $U(\B r_1, \B r_2, \cdots \B r_n)$ where $\B r_i$ is the position of the $i$'th particle. Generalization to 3-dimensional systems is straightforward if somewhat technical. The fundamental plastic instability  is most cleanly described in athermal ($T=0$) and quasi-static (AQS) conditions when an amorphous solid is subjected to quasi-static strain, allowing the system to regain mechanical equilibrium after every differential strain increase. Higher temperatures and finite strain rates introduce fluctuations and lack of mechanical equilibrium which cloud the fundamental physics
of plastic instabilities with unnecessary details \cite{10HKLP}.

In our AQS numerical simulations we
 use a $50-50$ binary Lennard-Jones mixture to simulate the shear localization discussed in this work. The potential energy for a pair of particles labeled  $i$ and $j$ has the form
\begin{eqnarray}
U_{ij}(r_{ij}) &=& 4\epsilon_{ij}\Big[\Big(\frac{\sigma_{ij}}{r_{ij}}\Big)^{12} - \Big(\frac{\sigma_{ij}}{r_{ij}}\Big)^{6} + A_0 \nonumber\\ &+& A_1\Big(\frac{r_{ij}}{\sigma_{ij}}\Big) + A_2\Big(\frac{r_{ij}}{\sigma_{ij}}\Big)^2\Big] \ , \label{Uij}
\end{eqnarray}
where the parameters $A_0$, $A_1$ and $A_2$ are added to smooth the potential at a scaled cut-off of $r/\sigma = 2.5$ (up to the second derivative). The parameters $\sigma_{AA}$, $\sigma_{BB}$ and $\sigma_{AB}$ were chosen to be $2\sin(\pi/10)$, $2\sin(\pi/5)$ and $1$ respectively and $\epsilon_{AA} = \epsilon_{BB} = 0.5, \epsilon_{AB} = 1$(see \cite{Langer1998}). The particle masses were taken to be equal. The samples were  prepared using high-temperature equilibration followed by a quench to zero temperature ($T=0.001$) (see \cite{Falk2007}).
For shearing, the usual athermal-quasistatic shear protocol was followed where each step comprises of an affine shift followed by an non-affine displacement using conjugate gradient minimization. The simulations were conducted in two dimensions (2d) and employed Lees-Edwards periodic boundary conditions. This implies that a square sample of size $L^2$ remains so also after strain. Samples were generated  with quench rates ranging from $3.2\times10^{-6}$ to $3.2\times 10^{-2}$ ( in LJ units ), and were strained to greater than $100$ percent. Simulations were performed on system-sizes ranging from $5000$ to $20000$ particles with a fixed density of $\rho = 0.976$ (in LJ units). The simulations reported in the paper have $10000$ particles and a quench-rate of $6.4\times 10^{-6}$ (in LJ units).

\begin{figure}
\includegraphics[scale = 0.20]{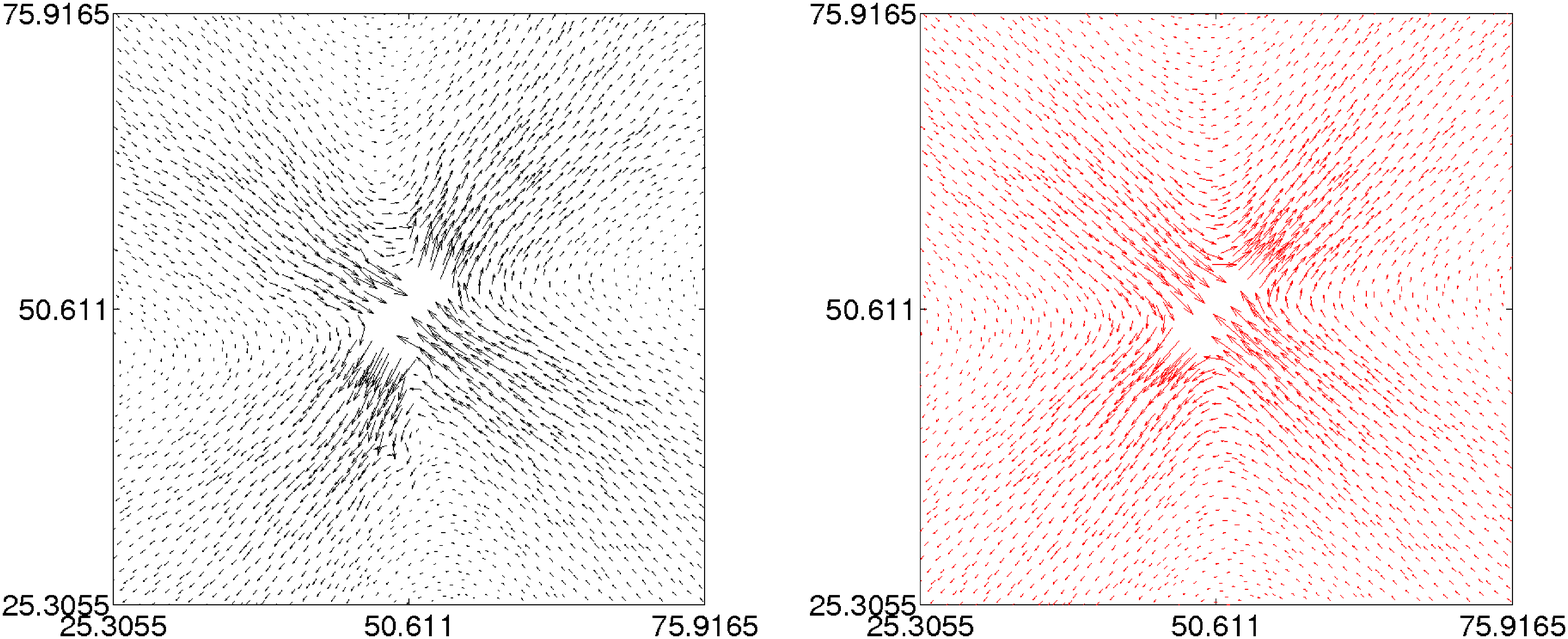}
\caption{(Color Online). Left panel: the localization of the non-affine displacement onto a quadrupolar structure which is modeled by an Eshelby inclusion, see right panel.  Right panel: the displacement field associated with a single Eshelby circular inclusion of radius $a$, see text.  The best fit parameters are $a \approx 2.5$ and $\epsilon^{*} \approx 0.1$. To remove the effect of boundary conditions, the best fit is generated on a smaller box of size $\left(x,y\right) \in \left[25.30 , 75.92\right]$ }
\label{loc}
\end{figure}

We choose to develop the theory for the case of external simple shear since then the strain tensor is traceless, simplifying some of the theoretical expressions. Applying an external shear, one discovers that the response of an amorphous solids to a small increase in the external shear strain $\delta\gamma$ (we drop tensorial indices for simplicity) is composed of two contributions. The first is the affine response which
simply follows the imposed shear, such that the particles positions $\B r_i={x_i,y_i}$ change via
\begin{eqnarray}
x_i &\to& x_i+\delta\gamma \,y_i \equiv x'_i\nonumber\\
y_i &\to& y_i \equiv y'_i.
\end{eqnarray}
 This affine response results in nonzero forces between the particles (in an amorphous solid) and these are relaxed by the non-affine response $\B u_i$ which returns
the system to mechanical equilibrium. Thus in total $\B r_i\to \B r'_i+\B u_i$. The nonaffine response $\B u_i$ solves an exact (and model independent) differential equation of the form \cite{ML,11HKLP}
\begin{equation}
\frac{d\B u_i}{d\gamma} = -H^{-1}_{ij} \Xi_j
\end{equation}
where $H_{ij} \equiv \frac{\partial^2 U(\B r_1, \B r_2, \cdots \B r_n)}{\partial \B r_i\partial \B r_j}$ is the so-called Hessian matrix and $\Xi_i\equiv
 \frac{\partial^2 U(\B r_1, \B r_2, \cdots \B r_n)}{\partial \gamma \partial \B r_i}$ is known as the non-affine force.
 The inverse of the Hessian matrix is evaluated after the removal of any Goldstone modes (if they exist). A plastic event occurs when a nonzero eigenvalue $\lambda_P$ of $\B H$ tends to zero at some strain value $\gamma_P$. It was proven that this occurs universally via a saddle node bifurcation such that $\lambda_P$ tends to zero like $\lambda_P\sim \sqrt{\gamma_P-\gamma}$ \cite{12DKP}. For values of the stress which are below the yield stress the plastic instability is seen \cite{ML} as
 a localization of the eigenfunction of $\B H$ denoted as $\Psi_P$ which is associated with the eigenvalue $\lambda_P$, (see Fig. \ref{loc} left panel).
 While at $\gamma=0$ all the eigenfunctions associated with low-lying eigenvalues are delocalized, $\Psi_P$ localizes as $\gamma\to \gamma_P$
 (when $\lambda_P\to 0$) on a quadrupolar structure as seen in Fig. \ref{loc} left panel for the non-affine displacement field when the plastic instability is approached. These simple plastic instabilities involve the motion of a relatively small number of particles (say 20 to 30 particles) but the stress field that is released has a long tail.

\begin{figure}
\includegraphics[scale = 0.19]{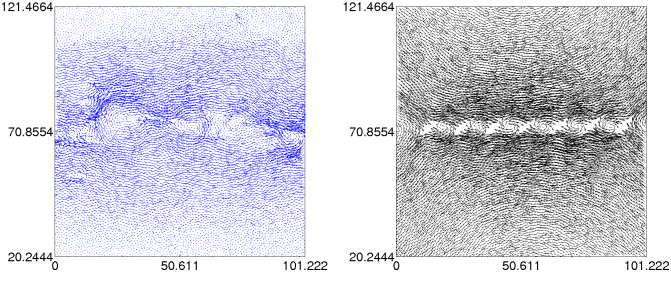}
\caption{(Color Online). Left panel: The nonaffine displacement field associated with a plastic instability that results in a shear band. Right panel: the displacement field associated with 7 Eshelby inclusions on a line with equal orientation. Note that in the left panel the quadrupoles are not precisely on a line as a result of the finite boundary conditions and the randomness. In the right panel the series of ${\C N}=7$ Eshelby inclusions, each given by Eq. (\ref{uc}) and separated by a distance of $13.158$, using the best fit parameters of Fig. \ref{loc}, have been superimposed to generate the displacement field shown. }
\label{result}
\end{figure}

 When the strain increases beyond some yield strain, the nature of the plastic instabilities can change in a fundamental way \cite{06TLB}. The main analytic calculation that is reported in Sect. \ref{energy}  shows that {\em when the stress built in the system is sufficiently large},
 instead of the eigenfunction localizing on a single quadrupolar structure, {\bf it can now localize on a series of $\C N$ such structures, which are organized on a line that is at $45$ degrees to the principal stress axis, with the quadrupolar structures having a fixed orientation relative to the applied shear}.  Fig. \ref{result} shows the non-affine field that is identical to the eigenfunction which is associated with this instability, clearly demonstrating the series
 of quadrupolar structures that are now organizing the flow such as to localize the shear in a narrow strip around them. This is the fundamental shear banding instability.
Note that this instability is reminiscent of some chainlike structure seen in liquid crystals, arising from the orientational elastic energy of the anisotropic host fluid \cite{97PSLW}, and ferromagnetic
chains of particles in strong magnetic fields \cite{70GP}. The reader should note that the event shown in Fig. \ref{result} will move the particles only a tiny amount, and it is the repeated
instability where many such event hit at the same region which results in the catastrophic
event that is seen as a shear band in experiments. Nevertheless this is the fundamental shear localization instability and in the sequel we will have to understand why repeated instabilities hit again and again in the same region. We should stress that this is not inevitable, for samples
that are prepared by a fast quench these instabilities appear at random places adding up to what seems to be a homogeneous flow. For further discussion of this point see Sect. \ref{summary}.

\section{Displacement  in 2d for a Circular Inclusion}
\label{inclusion}

 As said above, the shear localizing instability appears only when the stress exceeds a threshold.
 To explain why, we turn now to analysis. As a first step we model the quadrupolar stress field which is associated with the simple plastic instability as a circular Eshelby inclusion  \cite{ML}.
 \subsection{Circular Inclusion}
 \label{circular}
  We consider a 2d circular inclusion that has been strained into an ellipse using an eigenstrain or a stress-free strain $\epsilon _{\alpha\beta}^*$  which we take to be  traceless ie $\epsilon_{\gamma\gamma}^{*} = 0$ \cite{57Esh}. Here and below repeated indices imply a summation in 2-dimensions. A general expression for such a traceless tensor can be written in terms of a unit vector $\hat n_{\alpha}$ and a scalar $\epsilon^{*}$ as
\begin{eqnarray}\label{1}
\epsilon _{\alpha\beta}^{*} = \epsilon^{*}\left(2\hat n_{\alpha}\hat n_{\beta} - \delta_{\alpha\beta}\right) .
\end{eqnarray}
We also assume that a homogenous strain $\epsilon^{\infty}_{\alpha\beta}$ acts globally (which in our case  also triggers the local transformation of the inclusion). This strained ellipsoidal inclusion both feels a traction exerted by the surrounding elastic medium resulting in a constrained strain $\epsilon^{c}_{\alpha\beta}$ in the inclusion, and itself exerts a traction at the inclusion-elastic medium interface resulting in the originally unstrained surroundings developing a constrained strain field $\epsilon^{c}_{\alpha\beta}(\vec{X})$. Here and below
$\vec{X}$ stands for an arbitrary cartesian point in the material which for this purpose
is approximated as a continuum.

A fourth-order Eshelby tensor $S_{\alpha\beta\gamma\delta}$ can be defined which relates the constrained strain \textit{in} the inclusion $\epsilon^{c}_{\alpha\beta}$ to the eigenstrain $\epsilon _{\alpha\beta}^*$ viz.
\begin{eqnarray}\label{2}
\epsilon^{c}_{\alpha\beta} = S_{\alpha\beta\gamma\delta}\epsilon^{*}_{\gamma\delta} .
\end{eqnarray}
Now for an inclusion of arbitrary shape the constrained strain $\epsilon^{c}_{\alpha\beta}$, stress $\sigma^{c}_{\alpha\beta}$, and displacement field $u_{\alpha}$  inside the inclusion are in general functions of space. For ellipsoidal inclusions, however,  it was  shown by Eshelby \cite{E57,E59,2005LecNotes} that the Eshelby tensor and the constrained stress and strain fields \textit{inside} the inclusion become independent of space. We work here with a circular inclusion which is a special case of an ellipse and hence for such an inclusion, the Eshelby tensor is \cite{E57,E59,2005LecNotes}
\begin{eqnarray}\label{3}
S_{\alpha\beta\gamma\delta} = \frac{4\nu-1}{8\left(1-\nu\right)}\delta_{\alpha\beta}\delta_{\gamma\delta} + \frac{3-4\nu}{8\left(1-\nu\right)}\left(\delta_{\alpha\delta}\delta_{\beta\gamma} + \delta_{\beta\delta}\delta_{\alpha\gamma}\right) ,
\end{eqnarray}
where $\nu$ is the Poisson's ratio.  Note that this is the Eshelby tensor for an inclusion in 2-dimensions. It is the same as a cylindrical inclusion in 3-dimensions under a plane strain \cite{2005LecNotes}. From Eqs.~(\ref{2}) and (\ref{3}), we obtain
\begin{eqnarray}\label{5a}
\epsilon^{c}_{\alpha\beta} &=& \bigg[ \frac{4\nu-1}{8\left(1-\nu\right)}\delta_{\alpha\beta}\delta_{\gamma\delta} + \frac{3-4\nu}{8\left(1-\nu\right)}\left(\delta_{\alpha\delta}\delta_{\beta\gamma} + \delta_{\beta\delta}\delta_{\alpha\gamma}\right)\bigg]\epsilon^{*}_{\gamma\delta} \nonumber \\
&=& \frac{3-4\nu}{4\left(1-\nu\right)}\epsilon^{*}_{\alpha\beta} \quad \text{For a traceless eigenstrain} .
\end{eqnarray}
The total stress, strain and displacement field inside the circular inclusion is then given by
\begin{equation}\label{5b}
  \addtolength{\fboxsep}{5pt}
   \begin{gathered}
\epsilon^{I}_{\alpha\beta} = \epsilon^{c}_{\alpha\beta} + \epsilon^{\infty}_{\alpha\beta}  = \frac{3-4\nu}{4\left(1-\nu\right)}\epsilon^{*}_{\alpha\beta} + \epsilon^{\infty}_{\alpha\beta} \\
\sigma^{I}_{\alpha\beta} = \sigma^{c}_{\alpha\beta} - \sigma^{*}_{\alpha\beta} + \sigma^{\infty}_{\alpha\beta} \equiv C_{\alpha\beta\gamma\delta}\left(\epsilon^{c}_{\gamma\delta} - \epsilon^{*}_{\gamma\delta} + \epsilon^{\infty}_{\gamma\delta} \right) \\
u^{I}_{\alpha} = u^{c}_{\alpha} + u^{\infty}_{\alpha} = \left[\frac{3-4\nu}{4\left(1-\nu\right)}\epsilon^{*}_{\alpha\beta} + \epsilon^{\infty}_{\alpha\beta}\right]X_{\beta} \ .
   \end{gathered}
\end{equation}
Here the super-script $I$ indicates the inclusion and the $\sigma^{*}_{\alpha\beta}$ denotes the eigenstress which is linearly related to the eigenstrain by $\sigma^{*}_{\alpha\beta} \equiv C_{\alpha\beta\gamma\delta}\epsilon^{*}_{\gamma\delta}$, and which for an isotropic elastic medium simplifies further using
\begin{equation}
C_{\alpha\beta\gamma\delta} \equiv \lambda\delta_{\alpha\beta}\delta_{\gamma\delta} + \mu\left(\delta_{\alpha\gamma}\delta_{\beta\delta} + \delta_{\alpha\delta}\delta_{\beta\gamma}\right)\ , \label{calbe}
\end{equation}
to:
\begin{eqnarray}\label{6}
\sigma^{*}_{\alpha\beta} &=& 2\mu\epsilon^{*}_{\alpha\beta} + \lambda\epsilon^{*}_{\eta\eta}\delta_{\alpha\beta}\\ &=& \frac{\C E}{1+\nu}\epsilon^{*}_{\alpha\beta} + \frac{\C E\nu}{\left(1+\nu\right)\left(1-2\nu\right)}\epsilon^{*}_{\eta\eta}\delta_{\alpha\beta}\ ,\nonumber
\end{eqnarray}
where $\lambda$ and $\mu$ are the Lame's parameters. One can either choose the two Lame's coefficients or $\C E$ and $\nu$ as the two independent material parameters. The relations between them are given by
\begin{eqnarray}\label{7}
& \mu = \frac{\C E}{2\left(1+\nu\right)}\ , \quad
\lambda = \frac{\C E\nu}{\left(1+\nu\right)\left(1-2\nu\right)} \\
& \C E = \frac{\mu\left(3\lambda+2\mu\right)}{\lambda+\mu} \ ,
\quad \nu = \frac{\lambda}{2\left(\lambda+\mu\right)}
\end{eqnarray}
These relations are correct in 3-dimensions with plane strain conditions and therefore
also in 2-dimensions \cite{2005LecNotes}.

The stress in the inclusion can now be written down in terms of independent variables using Eq. (\ref{6}) by
\begin{eqnarray}\label{8}
\sigma^{I}_{\alpha\beta} &=& C_{\alpha\beta\gamma\delta}\left(\epsilon^{c}_{\gamma\delta} - \epsilon^{*}_{\gamma\delta} + \epsilon^{\infty}_{\gamma\delta} \right) \nonumber \\
&=& \frac{\C E}{1+\nu}\epsilon^{c}_{\alpha\beta}  + \frac{\C E\nu}{\left(1+\nu\right)\left(1-2\nu\right)}\epsilon^{c}_{\eta\eta}\delta_{\alpha\beta} \nonumber \\ &-& \frac{\C E}{1+\nu}\epsilon^{*}_{\alpha\beta} + \frac{\C E}{1+\nu}\epsilon^{\infty}_{\alpha\beta}
\end{eqnarray}
as $\epsilon^{*}_{\gamma\delta}$ and $\epsilon^{\infty}_{\gamma\delta}$ are traceless. Note that Eq. (\ref{5a}) implies that for a traceless eigenstrain, the constrained strain inside the  inclusion viz. $\epsilon^{c}_{\alpha\beta}$ is also
traceless and thus we obtain from Eq. (\ref{8}),
\begin{eqnarray}\label{10}
\sigma^{I}_{\alpha\beta} &=& \frac{\C E}{1+\nu}\epsilon^{c}_{\alpha\beta}  - \frac{\C E}{1+\nu}\epsilon^{*}_{\alpha\beta} + \frac{\C E}{1+\nu}\epsilon^{\infty}_{\alpha\beta} \nonumber \\
&=& \frac{\C E}{1+\nu}\frac{3-4\nu}{4\left(1-\nu\right)}\epsilon^{*}_{\alpha\beta}  - \frac{\C E}{1+\nu}\epsilon^{*}_{\alpha\beta} + \frac{\C E}{1+\nu}\epsilon^{\infty}_{\alpha\beta} \nonumber \\
&=& \frac{-\C E}{4\left(1+\nu\right)\left(1-\nu\right)}\epsilon^{*}_{\alpha\beta}  + \frac{\C E}{1+\nu}\epsilon^{\infty}_{\alpha\beta}
\end{eqnarray}

\subsection{Constrained Fields in the Elastic Medium}
\label{constrained}
\noindent In the surrounding elastic medium the stress, strain and displacement fields are all explicit function of space and can be written
\begin{equation}\label{epsm}
  \addtolength{\fboxsep}{5pt}
   \begin{gathered}
\epsilon^{m}_{\alpha\beta}(\vec{X}) = \epsilon^{c}_{\alpha\beta} (\vec{X}) + \epsilon^{\infty}_{\alpha\beta} \\
\sigma^{m}_{\alpha\beta} (\vec{X})= \sigma^{c}_{\alpha\beta} (\vec{X}) + \sigma^{\infty}_{\alpha\beta}  \\
u^{m}_{\alpha}(\vec{X}) = u^{c}_{\alpha}(\vec{X})+ u^{\infty}_{\alpha}(\vec{X}) \ .
   \end{gathered}
\end{equation}
In order to compute the displacement field $u^{c}_{\alpha}(\vec{X})$ in the isotropic elastic medium we need to solve the Lame-Navier equation
\begin{eqnarray}\label{12}
\frac{\C E}{2\left(1+\nu\right)\left(1-2\nu\right)}\frac{\partial^2 u^{c}_{\gamma}}{\partial X_{\alpha}\partial X_{\gamma}}\! + \!\!\frac{\C E}{2\left(1+\nu\right)}\frac{\partial^2 u^{c}_{\alpha}}{\partial X_{\beta} \partial X_{\beta}} = 0,
\end{eqnarray}
as there are no body forces present in our calculation. The constrained fields in the inclusion will supply the boundary conditions for the fields in the elastic medium at the inclusion boundary. Also as $r\rightarrow \infty$ the constrained displacement field will vanish.

All solutions of  equation Eq.~(\ref{12}) also obey the higher order bi-harmonic equation
\begin{eqnarray}\label{13}
\frac{\partial^4 u^{c}_{\alpha}}{\partial X_{\beta} \partial X_{\beta} \partial X_{\psi} \partial X_{\psi}} = \nabla^2 \nabla^2 u^c_{\alpha} = 0 .
\end{eqnarray}
Thus our objective is to  construct from the  radial solutions of the bi-laplacian equation Eq.~(\ref{13}) derivatives which also satisfy Eq.~(\ref{12}). Note that Eq.~(\ref{13}) is only a necessary (but not a sufficient) condition for the solutions and Eq.~(\ref{12}) still needs to be satisfied. The calculation is presented in Appendix A, with the final result
\begin{eqnarray}
&&u_\alpha^c (\vec X) =\label{uc}\\&&\frac{\epsilon^*}{4(1-\nu)}\left(\frac{a}{r}\right)^2\Big[2(1-2\nu) +\left(\frac{a}{r}\right)^2\Big]\Big[2\hat n_\alpha\B n\cdot\ \vec X- X_\alpha\Big]\nonumber\\&&+\frac{\epsilon^*}{2(1-\nu)}\left(\frac{a}{r}\right)^2
\left[1-\Big(\frac{a}{r}\right)^2\Big]\Big[\frac{2(\B n\cdot\vec X)^2}{r^2} -1\Big]X_\alpha \ . \nonumber
\end{eqnarray}
\subsection{Fit to the data}
Armed with this analytic expression we return now to our numerics, cf. Fig. \ref{loc}, and fit the
two parameters in Eq. (\ref{uc}) to the data of the displacement exhibited by a single localized
plastic event. The result of this procedure is $a\approx 2.5$ and $\epsilon^*\approx 0.1$. The quality of this
fit can be judged from the right panel of Fig.~\ref{loc} where we exhibit the form of Eq. (\ref{uc}) with the parameters fitted to the displacement field in the left panel. Also the value
of $a$ appears reasonable since it means that about $\pi a^2\approx 20$ particles are involved
in the core of the relaxation event. On physical grounds this is about the right order of magnitude.

We will keep these parameters fixed in all our calculations below. The reader should note that this is an approximation when there are multiple quadrupoles in the system, since they influence each other and the solution leading to Eq.~(\ref{uc}) should be repeated in the presence of many quadrupoles. We expect however
that the changes in the parameters should not be large when the density of the quadrupoles is small. We will always work in the small density limit $\rho a^2\ll 1$ where $\rho$ is the area
density of quadrupoles $\C N/L^2$.

\section{The energy of $\C N$ Eshelby Inclusions Embedded in a Matrix}
\label{energy}
\subsection{Notation}

Having the form of a single quadrupole, Eq. (\ref{uc}), we turn now to the calculation of the
energy associated with $\C N$ quadrupoles embedded in an elastic matrix. Once computed, we will show later that for large strains the minimum of this energy is obtained for a line of quadrupoles all having the same orientation.
From now on we use the notation that $u_{\alpha,\beta} \equiv \frac{\partial u_{\alpha}}{\partial \beta}$. The energy of $\C N$ Eshelby inclusions embedded in a linear elastic medium (or matrix) $\C N$, is given by the expression
\begin{eqnarray}\label{etot}
E\!\! = \!\!\frac{1}{2}\sum_{i=1}^{\C N}\int_{V_{0}^{(i)}}\!\!\sigma^{(i)}_{\alpha\beta}\epsilon^{(i)}_{\alpha\beta}dV \!+ \! \frac{1}{2}\int_{V -  \sum_{i=1}^{N}V_0^{(i)}}\!\!\sigma^{(m)}_{\alpha\beta}\epsilon^{(m)}_{\alpha\beta}dV
\end{eqnarray}
where the superscript $i$ indicates the index of the inclusion and $m$ indicates the matrix.
We evaluate Eq.~(\ref{etot}) in Appendix B. The result can be expressed in terms of four contributions: $E_{\rm mat}$ which is the contribution of the strained matrix, $E^\infty$ which is the energy of the $\C N$ qudrupoles in the external strain, $E_{\rm esh}$ which represents the self energy of the $\C N$ quadrupoles (their cost of creation) and lastly $E_{\rm inc}$ represents the
energy of interaction between the inclusions. Explicitly
\begin{widetext}
\begin{eqnarray}
& E_{mat} &\equiv \frac{1}{2}\sigma^{(\infty)}_{\alpha\beta}\epsilon_{\beta\alpha}^{(\infty)}V = V \sigma^{(\infty)}_{xy} \epsilon^{(\infty)}_{xy}  = \frac{V \C E\gamma^2}{2\left(1+\nu\right)} \label{Emat}\\
& E^{\infty} &\equiv  - \frac{1}{2}\sigma^{(\infty)}_{\alpha\beta}\left(\sum_{i=1}^{\C N}\epsilon^{(*,i)}_{\beta\alpha}V_0^{(i)}\right) = -\pi a^2\sigma_{xy}^{(\infty)}\sum_{i=1}^{\C N}\epsilon_{yx}^{(*,i)} =  -\frac{\pi a^2 \C E\gamma\epsilon^*}{\left(1+\nu\right)} \sum_{i=1}^{\C N} \hat n_x^{(i)} \hat n_y^{(i)} \label{Einf} \\
& E_{esh} &\equiv  - \frac{1}{2}\sum_{i=1}^{\C N}\epsilon_{\beta\alpha}^{(*,i)}\sigma^{(c,i)}_{\alpha\beta}V_0^{(i)} + \frac{1}{2}\sum_{i=1}^{\C N}\epsilon^{(*,i)}_{\beta\alpha}\sigma_{\alpha\beta}^{(*,i)}V_0^{(i)} = \frac{\pi a^2}{2} \sum_{i=1}^{\C N} \left(\sigma^{(*,i)}_{\alpha\beta} - \sigma^{(c,i)}_{\alpha\beta}\right)\epsilon^{(*,i)}_{\beta\alpha} \label{Eesh} \\
& E_{inc} &\equiv - \frac{1}{2} \sum_{i=1}^{\C N} \epsilon^{(*,i)}V_0^{(i)} \left(\sum_{j\neq i}\sigma_{\alpha\beta}^{(c,j)}\left(R_{ij}\right)\right) = -\frac{\pi a^2}{2}\sum_{\langle ij\rangle}\left[\epsilon^{(*,i)}_{\beta\alpha}\sigma^{(c,j)}_{\alpha\beta}\left(R_{ij}\right) + \epsilon^{(*,j)}_{\beta\alpha}\sigma^{(c,i)}_{\alpha\beta}\left(R_{ij}\right)\right] \label{Einc}
\end{eqnarray}
The above expressions are specific to 2D, for a global strain corresponding to simple shear under the linear approximation. Thus $\epsilon^{\infty}_{xy} = \frac{\gamma}{2}$ here, and the traceless eigenstrain takes the form $\epsilon^{(*,i)}_{yx} = 2\epsilon^{*}n_x^{(i)}n_y^{(i)}$.\\

\begin{figure}
\begin{center}
\includegraphics[scale = 0.35]{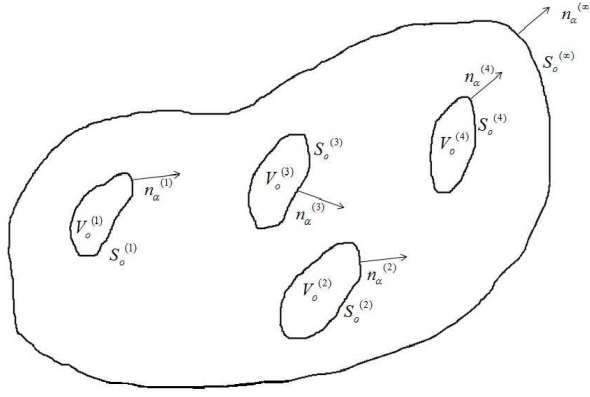}
\caption{Schematic of four Eshelby inclusions embedded in a matrix `m'}
\end{center}
\label{a}
\end{figure}
The form of $E_{\rm inc}$ is not final, and we bring it to its final form in Appendix C. The
final result is

\begin{eqnarray}\label{Esubinc}
E_{inc} &=& -\frac{\C E(\epsilon^*)^2\pi a^2}{8(1-\nu^2)}\sum_{\langle ij\rangle}\left(\frac{a}{R_{ij}}\right)^2 \times \nonumber \\ && \bigg[-8\left\lbrace\left(1-2\nu\right) + \left(\frac{a}{R_{ij}}\right)^2\right\rbrace \bigg(4\left(\mathbf{\hat n}^{(i)}\cdot \mathbf{\hat n}^{(j)}\right) \left(\mathbf{\hat n}^{(i)}\cdot \mathbf{\hat r}_{ij}\right) \left(\mathbf{\hat n}^{(j)}\cdot \mathbf{\hat r}_{ij}\right) \nonumber \\
&-& 2\left(\mathbf{\hat n}^{(i)} \cdot \mathbf{\hat r}_{ij}\right)^2 - 2\left(\mathbf{\hat n}^{(j)} \cdot \mathbf{\hat r}_{ij}\right)^2 + 1\bigg) + 4\left(2\left(1-2\nu\right) + \left(\frac{a}{R_{ij}}\right)^2\right)\left(2\left(\mathbf{\hat n}^{(i)}\cdot \mathbf{\hat n}^{(j)}\right)^2 - 1\right) \nonumber \\
&-&8 \left( 1 - 2\left(\frac{a}{R_{ij}}\right)^2\right) \left( 2\left(\mathbf{\hat n}^{(i)} \cdot \mathbf{\hat r}_{ij}\right)^2 - 1\right)\left(2\left(\mathbf{\hat n}^{(j)} \cdot \mathbf{\hat r}_{ij}\right)^2 - 1\right) \nonumber \\
&+& 32\left(1 - \left(\frac{a}{R_{ij}}\right)^2\right) \left(\left(\mathbf{\hat n}^{(i)}\cdot \mathbf{\hat r}_{ij}\right)\left(\mathbf{\hat n}^{(j)}\cdot \mathbf{\hat r}_{ij}\right)\left(\mathbf{\hat n}^{(i)} \cdot \mathbf{\hat n}^{(j)}\right) - \left(\mathbf{\hat n}^{(i)}\cdot \mathbf{\hat r}_{ij}\right)^2\left(\mathbf{\hat n}^{(j)}\cdot \mathbf{\hat r}_{ij}\right)^2\right)\bigg]
\end{eqnarray}
\end{widetext}
where $\mathbf{\hat r}_{ij} \equiv \frac{\vec X^{ij}}{R_{ij}}$.

Our task is now to find the configuration of $\C N$ Eshelby quadrupoles that minimize the total energy. Obviously, if the external strain $\gamma$ is sufficiently large, we need to minimize
$E^\infty$ separately, since it is proportional to $\gamma$. The minimum of (\ref{Einf}) is
obtained for
\begin{equation}
n_x^{(i)}= n_y^{(i)} = \frac{1}{\sqrt{2}} \ . \label{nxny}
\end{equation}
Substituting this result in Eq. (\ref{Esubinc}) simplifies it considerably. We find
\begin{widetext}
\begin{eqnarray}
&&E_{\rm inc} = - \frac{\pi a^2(\epsilon^*)^2 \C E}{8(1-\nu^2)} \sum_{<ij>} (\frac{a}{R_{ij}})^2\Big\{
-8[(1-2\nu)+(\frac{a}{R_{ij}})^2]+4[2(1-2\nu) + (\frac{a}{R_{ij}})^2]
-8[1-2(\frac{a}{R_{ij}})^2][2(\hat {\B n} \cdot \hat {\B r})^2-1]^2\nonumber\\
&&+32[1-(\frac{a}{R_{ij}})^2][(\hat {\B n} \cdot \hat {\B r})^2-(\hat {\B n} \cdot \hat {\B r})^4]
\Big\} \ . \label{final}
\end{eqnarray}
\end{widetext}
We can find the minimum energy very easily. Denote $x\equiv (\hat {\B n} \cdot \hat {\B r})^2$, and minimize the expression $A[2x-1]^2 - B[x-x^2]$. The minimum is obtained
at $x=1/2$, or $\cos \phi = \sqrt{1/2}$. We thus conclude that when the line of correlated quadrupole forms under shear,  this line is in 45 degrees to the compressive axis, as is indeed
seen in experiments. Of course there are two solutions for this, perpendicular to each other. Note that for other external strains which are not consistent with a traceless
strain tensor (or in 3-dimensions) this conclusion may change.

The physical meaning of this analytic result is that it is cheaper (in energy) for the material
to organize $\C N$ quadrupolar structures on a line of 45 degrees with the compressive stress, all having the same orientation, than any other arrangement of these $\C N$ quadrupoles, including any
random distribution. This explains why such a highly correlated distribution appears in the strained amorphous solid, and why it can only appear when the external strain (or the built-up stress) are high enough. This fact, in addition to the observation that such an arrangement of Eshelby quadrupoles organizes the displacement field into a localized shear, explains the origin of this fundamental instability.

\section{Estimate of yield-stress and number of Eshelby quadrupoles}
\label{estimate}
In this section we turn to estimate the yield stress and the associated density of
Eshelby quadrupoles. To this aim we need to compute one other energy term that was
not needed until now, namely $E_{\rm esh}$ which was the same for all the configurations
of the quadrupoles.
\subsection{Expression for $E_{esh}$}
The energy term $E_{\rm esh}$ was given by Eq. (\ref{Eesh}) which
can be re-written as
\begin{eqnarray}\label{34}
&=& \frac{\pi a^2}{2}\sum_{i=1}^{\C N} \epsilon^{(*,i)}_{\beta\alpha} C_{\alpha\beta\gamma\delta}\left(\epsilon^{(*,i)}_{\gamma\delta} - \epsilon^{(c,i)}_{\gamma\delta}\right) \\
&=& \frac{\pi a^2}{2}\sum_{i=1}^{\C N} \epsilon^{(*,i)}_{\beta\alpha} C_{\alpha\beta\gamma\delta}\left(\epsilon^{(*,i)}_{\gamma\delta} - \frac{3- 4\nu}{4(1-\nu)}\epsilon^{(*,i)}_{\gamma\delta} \right) \ , \   \text{Cf.  Eq. \ref{5a}} \nonumber \\
&=& \frac{\pi a^2}{2}\sum_{i=1}^{\C N} \epsilon^{(*,i)}_{\beta\alpha} C_{\alpha\beta\gamma\delta} \frac{\epsilon^{(*,i)}_{\gamma\delta}}{4(1-\nu)}\ .\nonumber
\end{eqnarray}
For an isotropic matrix, using Eq. (\ref{calbe}),  we have
\begin{eqnarray}\label{35}
C_{\alpha\beta\gamma\delta} \epsilon^{(*,i)}_{\gamma\delta} &=& \lambda\delta_{\alpha\beta}\delta_{\gamma\delta} \epsilon^{(*,i)}_{\gamma\delta} + \mu\left(\delta_{\alpha\gamma}\delta_{\beta\delta} + \delta_{\alpha\delta}\delta_{\beta\gamma}\right) \epsilon^{(*,i)}_{\gamma\delta} \nonumber \\
&=& 2\mu \epsilon^{(*,i)}_{\alpha\beta}  = \frac{\C E}{1+\nu}\epsilon^{(*,i)}_{\alpha\beta}
\end{eqnarray}
for a symmetric traceless eigenstrain. Using Eq. (\ref{35}) for circular inclusions each of radius $a$ in 2D, we obtain
\begin{eqnarray}\label{36}
E_{esh} = \frac{\pi a^2}{2} \frac{\C E}{4(1-\nu^2)} \sum_{i=1}^{\C N} \epsilon^{(*,i)}_{\beta\alpha}\epsilon^{(*,i)}_{\alpha\beta}
\end{eqnarray}
Now,
\begin{eqnarray}\label{37}
&&\epsilon^{(*,i)}_{\beta\alpha}\epsilon^{(*,i)}_{\alpha\beta} = (\epsilon^{(*,i)})^2\left(2\hat n^{(i)}_{\alpha}\hat n^{(i)}_{\beta} - \delta_{\alpha\beta}\right)\left(2\hat n^{(i)}_{\beta}\hat n^{(i)}_{\alpha} - \delta_{\beta\alpha}\right)\nonumber\\ &&= 2(\epsilon^{(*,i)})^2
\end{eqnarray}

For all eigenstrains equal (ie. $\epsilon^{(*,i)} = \epsilon^{*}$), we have
\begin{eqnarray}\label{38}
E_{esh} = \frac{\C E\pi a^2 \C N (\epsilon^{*})^2 }{4(1-\nu^2)}
\end{eqnarray}
\subsection{$E_{\rm inc}$ for a line of equidistant quadrupoles with the same polarization}

At this point we need to compute the energy term $E_{\rm inc}$ for the special configurations of $\C N$ quadrupoles that are equi-distant and with
the same polarization, organized in a line. In this case we have
\begin{eqnarray}\label{39}
\mathbf{\hat n^{(i)}} = \mathbf{\hat n}, \quad \mathbf{\hat r_{ij} = \hat r}, \texttt{and} \quad R_{ij} = |j-i|R
\end{eqnarray}
Starting from Eq. (\ref{final}) we specialize to the present situation
\begin{widetext}
\begin{eqnarray}\label{40}
E_{inc} &=& -\frac{\C E(\epsilon^*)^2\pi a^2}{8(1-\nu^2)}\sum_{i=1}^{\C N-1} \sum_{j, j > i}^{\C N}\left(\frac{a}{R}\right)^2\frac{1}{(j-i)^2} \Bigg\{ -8\left[\left(1-2\nu\right) + \left(\frac{a}{R}\right)^2\frac{1}{(j-i)^2}\right] \nonumber \\
&+& 4\left[2\left(1-2\nu\right) + \left(\frac{a}{R}\right)^2\frac{1}{(j-i)^2}\right]
-8 \left[ 1 - 2\left(\frac{a}{R}\right)^2\frac{1}{(j-i)^2}\right] \left[ 2\mathbf{\left(\hat n \cdot \hat r\right)}^2 - 1\right]^2 \nonumber \\
&+& 32\left[1 - \left(\frac{a}{R}\right)^2\frac{1}{(j-i)^2}\right]) \left[\mathbf{\left(\hat n \cdot \hat r\right)}^2 - \mathbf{\left(\hat n \cdot \hat r\right)}^4\right]\bigg\}\nonumber \\
\end{eqnarray}
For $(\mathbf{\hat n\cdot \hat r)}^2 = 1/2$, the above expression reduces to
\begin{eqnarray}\label{41}
E_{inc}= -\frac{\C E(\epsilon^*)^2\pi a^2}{2(1-\nu^2)}\sum_{i=1}^{\C N-1} \sum_{j, j > i}^{\C N}\bigg[\frac{2}{(j-i)^2}\left(\frac{a}{R}\right)^2 - 3\left(\frac{a}{R}\right)^4\frac{1}{(j-i)^4}\bigg]
\end{eqnarray}
We have
\begin{eqnarray}\label{42}
\sum_{i=1}^{\C N-1} \sum_{j = i+1}^{\C N}\frac{1}{(j-i)^s} = \sum_{i=1}^{\C N-1} \sum_{n=1}^{\C N-i}\frac{1}{n^s} \approx \C N\zeta (s) \quad \texttt{for} \quad \C N >> 1 \ , \
\end{eqnarray}
where $\zeta(s)$ is the Riemann zeta function. Thus we obtain
\begin{eqnarray}
E_{inc} = -\frac{\C E(\epsilon^*)^2\pi a^2 \C N} {2(1-\nu^2)}\left[2\left(\frac{a}{R}\right)^2\zeta(2) - 3\left(\frac{a}{R}\right)^4\zeta(4)\right] \ . \label{ER}
\end{eqnarray}
\end{widetext}

At this point we realize that the distance $R$ between the quadrupoles is not determined. We
will choose $R$ by demanding that the line density of the quadrupoles remains invariant
in the thermodynamic limit, or $R\equiv L/{\C N}$. Thus, with $\rho\equiv \C N/L$,  the energy density in the strip $L\times a$ reads
\begin{eqnarray}\label{43}
\frac{E_{inc}}{La} = -\frac{\C E(\epsilon^*)^2\pi a\rho}{2(1-\nu^2)}\bigg[2(a\rho)^2\zeta(2) - 3(a\rho)^4\zeta(4)\bigg]
\end{eqnarray}
Similarly from equation \ref{38}, we obtain
\begin{eqnarray}\label{44}
\frac{E_{esh}}{La} = \frac{\C E\pi a(\epsilon^{*})^2 }{4(1-\nu^2)}  \rho
\end{eqnarray}
From equation (\ref{nxny}) we have
\begin{eqnarray}\label{45}
\frac{E^{\infty}}{La}  = -\frac{\pi a\C E\gamma\epsilon^*}{2\left(1+\nu\right)} \rho
\end{eqnarray}
Thus the plastic energy density is given by
\begin{eqnarray}\label{46a}
&&\frac{E(\rho,\gamma)}{La} \equiv \frac{E^\infty+E_{\rm esh}+E_{\rm inc}}{La}
\\&&= \frac{\C E\pi (\epsilon^{*})^2}{4(1-\nu^2)}\bigg[A\left(1-\frac{\gamma}{\gamma_{_{\rm Y}}}\right)\rho a - B(\rho a)^3 + C(\rho a)^5\bigg] \ , \nonumber
\end{eqnarray}
where $A = 1, B= 4\zeta(2)$ and $C =6\zeta(4)$. $\gamma_{_{\rm Y}}$ is defined as
\begin{equation}
\gamma_{_{Y}} \equiv \frac{\epsilon^*}{2(1-\nu)}.
\end{equation}
\begin{figure}
\label{Eplast}
\includegraphics[scale = 0.15]{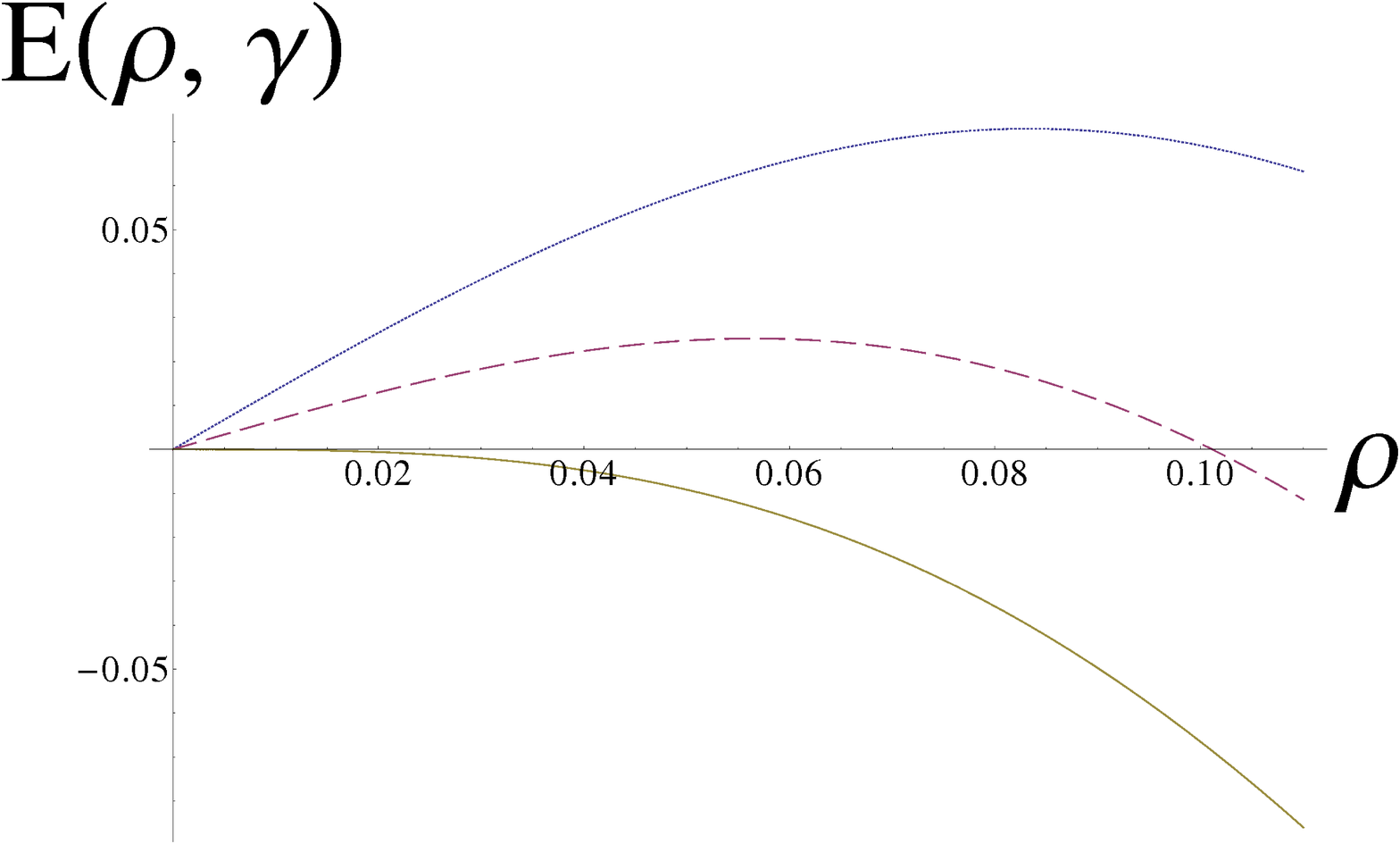}
\caption{(Color Online). The total plastic energy for the creation of an array of quadrupoles with line density
$\rho$ for three values of $\gamma$: $\gamma=\gamma_{_{\rm Y}}-0.1$,
$\gamma=\gamma_{_{\rm Y}}-0.05$,
and $\gamma=\gamma_{_{\rm Y}}$}
\end{figure}
Eq. (\ref{46a}) is plotted, using the numerical values of the parameters found in our
numerical simulations in Fig. \ref{Eplast} for various values of $\gamma$. For $\gamma<\gamma_{_{\rm Y}}$ the minimum of the expression is attained always at $\rho=0$, and
the shear localization cannot occur. Only at $\gamma\ge \gamma_{_{\rm Y}}$ a new solution
opens up to allow a finite density of the Eshelby quadrupoles. Therefore $\gamma_{_{\rm Y}}$
is by definition the yield strain.

\section{Summary and Conclusions}
\label{summary}

We have presented a theory of the fundamental instability that leads to shear localization and
eventually to shear bands. One remarkable observation is that the natural plastic instability that
occurs spontaneously in our simulations results in a displacement field that is surprisingly close
to the one made by an Eshelby circular inclusion, see Fig. \ref{loc}. The best fit for the parameter $a$, of the order of 2.5 is in agreement with the intuitive belief that shear transformation zones involve 20-30 particles, as $\pi a^2$ would predict. Basing our analysis on this similarity we could develop an analytic theory of the energy needed to create $\C N$ such
inclusions, whether scattered in the system randomly or aligned and organized in highly correlated
shear localized structure. We discover that the latter becomes energetically favorable when
$\gamma$ exceeds $\gamma_{_{Y}} \equiv \frac{\epsilon^*}{2(1-\nu)}$. In our system $\nu\approx 0.215$, and with our best fit $\epsilon^*\approx 0.1$ we predict $\gamma_{_{Y}}\approx 0.07$ which is right on the mark as one can seen from Ref. \cite{12DHP}.

While we believe that our calculation of the energy of $\C N$ quadrupoles is accurate for
densities such that $\rho a^2\ll 1$, the interaction between the quadrupoles become much
more involved for higher densities, and we avoided this complication. The consequence is that
we cannot predict a-priory the critical density of our
 quadrupoles, and we leave this interesting issue for future research. Another issue that
 warrants further study is whether the parameters $a$ and $\epsilon^*$ are material parameters
 which are determined by the small scale structure of the glass, and if so, how to estimate them
 a-priori. Also, are these parameters dependent on the way the system is stressed, i.e. via shear
 or via tensile compression etc..

\begin{figure}
\includegraphics[scale = 0.30]{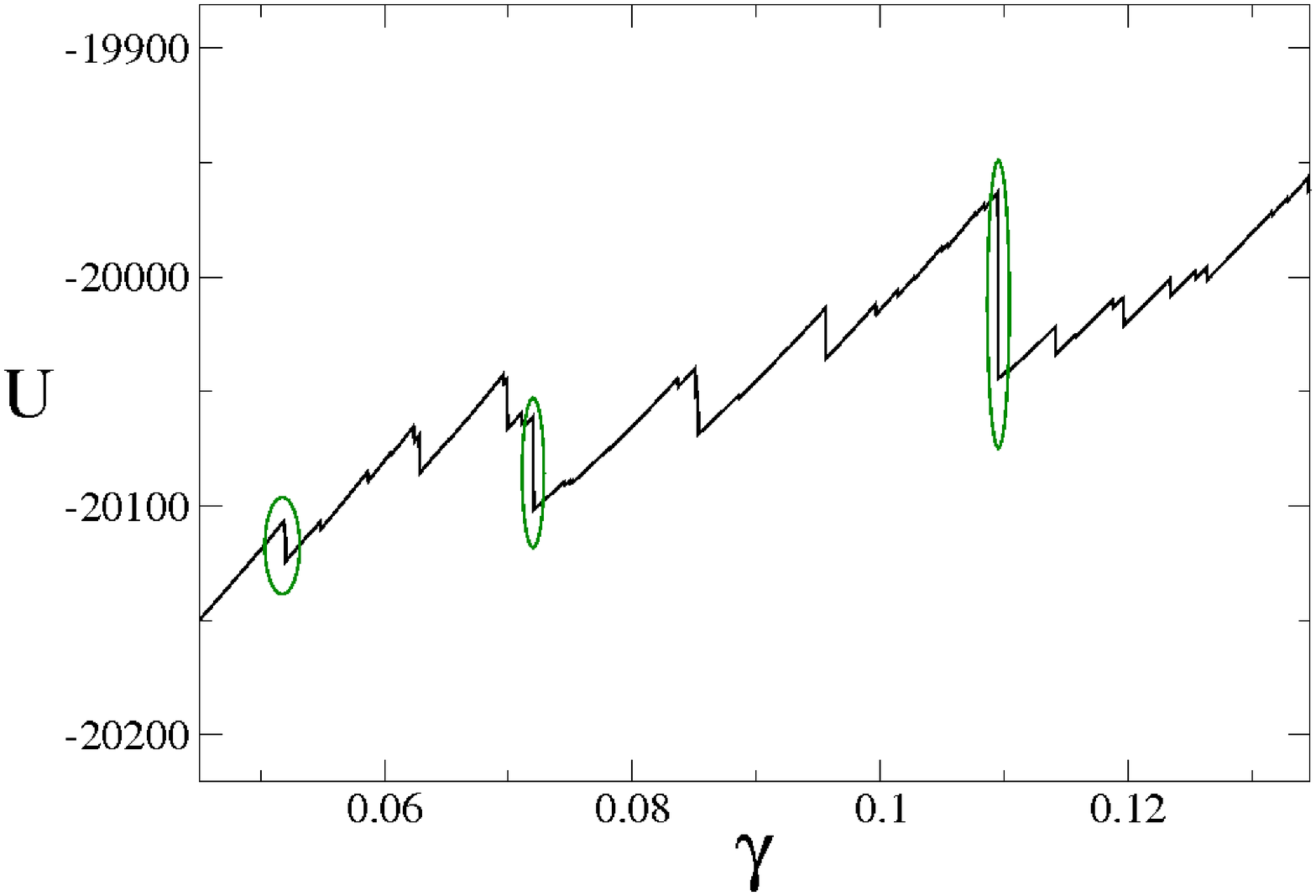}
\includegraphics[scale = 0.20]{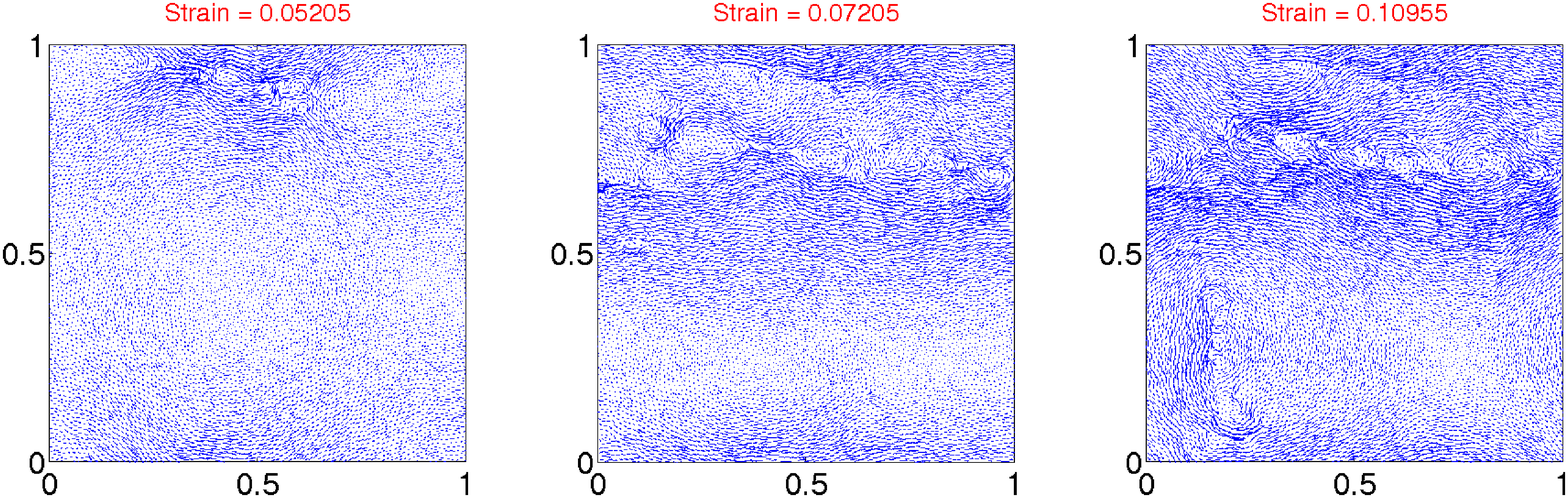}
\caption{(Color Online). Repeated shear localization instabilities in the case of slow quench. Upper panel: the
energy vs. strain and the instabilities that were chosen for display in the lower panel. The average non affine displacement field after the instabilities that are marked in the upper panel. The point to notice is that the instability falls repeatedly on the same band, accumulating to a shear band.}
\label{again}
\end{figure}

Next we need to discuss the difference in behavior of glasses that were quenched relatively
 quickly and those that were quenched relatively slowly. In our simulations we find that the latter exhibit the fundamental shear localization instability again and again along approximately the same line, accumulating displacement that develops into a shear band. On the other hand the former tends every time to have plastic instabilities at different places, sometime localized and sometime more correlated, and on the average this then appears like a homogenous flow. This difference is stressed in Fig. \ref{again} where the case of slow
 quench is exhibited. It appears that fastly
 quenched systems may have plastic instabilities almost everywhere, and there is no special preference for one line or another. Slowly quenched systems are initially harder to shear localize, but once it happens in a particular (random) line it is likely to repeat again and again along the same line. Making these words quantitative is again an issue left to future research. In particular it is interesting to find what is the precise meaning of fast and slow quenches, fast and slow compared to what, and how this changes the microscopic local structure.

 Finally the effects of temperature and finite strain rates on the present mechanism constitute
 a separate piece of work which of course is of the utmost importance. Like the other
 open subject mentioned above, it will be taken up in future research.

\acknowledgments
This work had been supported in part by the Israel Science Foundation, the German-Israeli Foundation and by the European Research Council under an ``ideas" grant.\\

\appendix

\section{The displacement field of an Eshelby circular inclusion}

\subsection{Solutions of the Lame-Navier equation}

\noindent We look for linear combinations of derivatives of the radial solutions of Eq. \ref{13} which are linear in the eigenstrain $\epsilon^{*}_{\alpha\beta}$ and go to zero at large radii. In addition the terms must transform as components of a vector field. Such a solution can be written down as
\begin{eqnarray}\label{14}
u^{c}_{\alpha} &=& A\epsilon^{*}_{\alpha\beta}\frac{\partial \ln r}{\partial X_{\beta}} + B \epsilon^{*}_{\beta\gamma}\frac{\partial^3 \ln r}{\partial X_{\alpha} \partial X_{\beta} \partial X_{\gamma}} \nonumber \\&+& C\epsilon^{*}_{\beta\gamma}\frac{\partial^3\left( r^2\ln r\right)}{\partial X_{\alpha} \partial X_{\beta} \partial X_{\gamma}} \ ,
\end{eqnarray}
where $X_\alpha$ is the $\alpha$ component of the position vector with the origin at the center
of the Eshelby quadrupole, and $r\equiv |\vec{\B X}|$.
It can be checked that any other terms are either zero,  do not go to zero as $r \rightarrow \infty$, or do not transform as components of a vector. We re-write equation Eq.~(\ref{12}) as
\begin{eqnarray}\label{16}
\left(\frac{1}{1-2\nu}\right)\frac{\partial^2 u^{c}_{\gamma}}{\partial X_{\alpha}\partial X_{\gamma}} + \frac{\partial^2 u^{c}_{\alpha}}{\partial X_{\beta} \partial X_{\beta}} = 0
\end{eqnarray}
which is the equation for the constrained displacement field in the elastic matrix subject to appropriate boundary conditions.
From Eq. (\ref{14}), we obtain
\begin{widetext}
\begin{eqnarray}\label{17}
\frac{\partial^2 u^{c}_{\alpha}}{\partial X_{\beta} \partial X_{\beta}}\! = \! A\epsilon^{*}_{\alpha\eta}\frac{\partial}{\partial X_{\eta}}\!\left[\frac{\partial^2}{\partial X_{\beta} \partial X_{\beta}}\ln r\right]\!\! + \!\!B \epsilon^{*}_{\eta\lambda}\frac{\partial^3}{\partial X_{\alpha} \partial X_{\eta} \partial X_{\lambda}}\!\left[\frac{\partial^2}{\partial X_{\beta} \partial X_{\beta}}\ln r\right] \!\!+\! C\epsilon^{*}_{\eta\lambda}\frac{\partial^3}{\partial X_{\alpha} \partial X_{\eta} \partial X_{\lambda}}\!\left[\frac{\partial^2}{\partial X_{\beta} \partial X_{\beta}}r^2\ln r\right] \ .
\end{eqnarray}
\end{widetext}
We need the following identities
\begin{eqnarray}\label{18}
\begin{gathered}
\frac{\partial^2}{\partial X_{\beta} \partial X_{\beta}}\left(\ln r\right) = 0  \\
\frac{\partial^2}{\partial X_{\beta} \partial X_{\beta}}\left( r^2\ln r\right) = 4\ln r +4
\end{gathered}
\end{eqnarray}
Thus we obtain from equation \ref{17},
\begin{eqnarray}\label{19}
\frac{\partial^2 u^{c}_{\alpha}}{\partial X_{\beta} \partial X_{\beta}} = 4C\epsilon^{*}_{\eta\lambda}\frac{\partial^3 \ln r}{\partial X_{\alpha} \partial X_{\eta} \partial X_{\lambda}}
\end{eqnarray}
Similarly, the expression for
\begin{widetext}
\begin{eqnarray}\label{20}
\frac{\partial^2 u^{c}_{\gamma}}{\partial X_{\alpha}\partial X_{\gamma}} &=& \frac{\partial^2}{\partial X_{\alpha}\partial X_{\gamma}}\left[A\epsilon^{*}_{\gamma\eta}\frac{\partial \ln r}{\partial X_{\eta}} + B \epsilon^{*}_{\eta\lambda}\frac{\partial^3 \ln r}{\partial X_{\gamma} \partial X_{\eta} \partial X_{\lambda}} + C\epsilon^{*}_{\eta\lambda}\frac{\partial^3\left( r^2\ln r\right)}{\partial X_{\gamma} \partial X_{\eta} \partial X_{\lambda}}\right] \nonumber \\
&=& \frac{\partial}{\partial X_{\alpha}}\left[A\epsilon^{*}_{\gamma\eta}\frac{\partial^2 \ln r}{\partial X_{\gamma}\partial X_{\eta}} + B \epsilon^{*}_{\eta\lambda}\frac{\partial^4 \ln r}{\partial X_{\gamma} \partial X_{\gamma} \partial X_{\eta} \partial X_{\lambda}} + C\epsilon^{*}_{\eta\lambda}\frac{\partial^4\left( r^2\ln r\right)}{\partial X_{\gamma} \partial X_{\gamma}\partial X_{\eta} \partial X_{\lambda}}\right] \nonumber \\
\end{eqnarray}
\end{widetext}
which can be re-written (noting that the second and third terms involve the laplacian for which we have identities from Eqs. (\ref{18}) as
\begin{eqnarray}\label{21}
\frac{\partial^2 u^{c}_{\gamma}}{\partial X_{\alpha}\partial X_{\gamma}} &=& \frac{\partial}{\partial X_{\alpha}}\left[A\epsilon^{*}_{\gamma\eta}\frac{\partial^2 \ln r}{\partial X_{\gamma}\partial X_{\eta}} + C\epsilon^{*}_{\eta\lambda}\frac{\partial^2\left( 4\ln r + 4\right)}{\partial X_{\eta} \partial X_{\lambda}}\right] \nonumber \\
&=& \left(A+4C\right)\epsilon^{*}_{\eta\lambda}\frac{\partial^3 \ln r}{\partial X_{\alpha} \partial X_{\eta} \partial X_{\lambda}}
\end{eqnarray}
Plugging expressions (\ref{19}) and (\ref{21}) in Eq. (\ref{16}), we thus obtain
\begin{widetext}
\begin{eqnarray}\label{22}
&\frac{\left(A+4C\right)}{1-2\nu}\epsilon^{*}_{\eta\lambda}\frac{\partial^3 \ln r}{\partial X_{\alpha} \partial X_{\eta} \partial X_{\lambda}} + 4C\epsilon^{*}_{\eta\lambda}\frac{\partial^3 \ln r}{\partial X_{\alpha} \partial X_{\eta} \partial X_{\lambda}} = 0  \Rightarrow \left[\frac{A+4C}{1-2\nu} + 4C\right]\epsilon^{*}_{\eta\lambda}\frac{\partial^3 \ln r}{\partial X_{\alpha} \partial X_{\eta} \partial X_{\lambda}} = 0 \Rightarrow C = -\frac{A}{8\left(1-\nu\right)}
\end{eqnarray}
We can thus re-write Eq. (\ref{14}) as
\begin{eqnarray}\label{23}
u^{c}_{\alpha} = A\epsilon^{*}_{\alpha\beta}\frac{\partial \ln r}{\partial X_{\beta}} + B \epsilon^{*}_{\beta\gamma}\frac{\partial^3 \ln r}{\partial X_{\alpha} \partial X_{\beta} \partial X_{\gamma}} -\frac{A}{8\left(1-\nu\right)}\epsilon^{*}_{\beta\gamma}\frac{\partial^3\left( r^2\ln r\right)}{\partial X_{\alpha} \partial X_{\beta} \partial X_{\gamma}}
\end{eqnarray}
The following identities are now required:
\begin{eqnarray}\label{24}
\frac{\partial \ln r}{\partial X_{\beta}} &=& \frac{X_{\beta}}{r^2} \nonumber \\
\frac{\partial^3 \ln r }{\partial X_{\alpha} \partial X_{\beta} \partial X_{\gamma}} &=& \frac{-2r^2\left(X_{\alpha}\delta_{\beta\gamma}+X_{\beta}\delta_{\alpha\gamma} + X_{\gamma}\delta_{\alpha\beta}\right) + 8X_{\alpha} X_{\beta} X_{\gamma}}{r^6} \nonumber \\
\frac{\partial^3\left(r^2\ln r\right)}{\partial X_{\alpha} \partial X_{\beta} \partial X_{\gamma}} &=& \frac{2r^2\left(X_{\alpha}\delta_{\beta\gamma} + X_{\beta}\delta_{\alpha\gamma} + X_{\gamma}\delta_{\alpha\beta}\right) - 4X_{\alpha} X_{\beta} X_{\gamma}}{r^4}
\end{eqnarray}
Using these relations we can re-write Eq.~\ref{23} as
\begin{eqnarray}\label{25}
u^{c}_{\alpha} &=& A\epsilon^{*}_{\alpha\beta} \frac{X_{\beta}}{r^2} + B \epsilon^{*}_{\beta\gamma}\left[\frac{-2r^2\left(X_{\alpha}\delta_{\beta\gamma}+X_{\beta}\delta_{\alpha\gamma} + X_{\gamma}\delta_{\alpha\beta}\right) + 8X_{\alpha} X_{\beta} X_{\gamma}}{r^6}\right] \nonumber\\ &-&\frac{A}{8\left(1-\nu\right)}\epsilon^{*}_{\beta\gamma}\left[\frac{2r^2\left(X_{\alpha}\delta_{\beta\gamma} + X_{\beta}\delta_{\alpha\gamma} + X_{\gamma}\delta_{\alpha\beta}\right) - 4X_{\alpha} X_{\beta} X_{\gamma}}{r^4}\right] \nonumber \\
&=& A\epsilon^{*}_{\alpha\beta} \frac{X_{\beta}}{r^2} -\left[\frac{2B}{r^4} + \frac{A}{4\left(1-\nu\right)r^2} \right]\epsilon^{*}_{\beta\gamma}\left(X_{\alpha}\delta_{\beta\gamma}+X_{\beta}\delta_{\alpha\gamma} + X_{\gamma}\delta_{\alpha\beta}\right)
  + \left[\frac{8B}{r^6} + \frac{A}{2\left(1-\nu\right)r^4}\right]\epsilon^{*}_{\beta\gamma}X_{\alpha} X_{\beta} X_{\gamma}
\end{eqnarray}
Remembering that the eigenstrain is traceless,we find that $\epsilon^{*}_{\beta\gamma}\left(X_{\alpha}\delta_{\beta\gamma}+X_{\beta}\delta_{\alpha\gamma} + X_{\gamma}\delta_{\alpha\beta}\right) = 2\epsilon^{*}_{\alpha\beta}X_{\beta}$ using which we can simplify Eq. (\ref{25}) to obtain
\begin{eqnarray}\label{26}
& u^{c}_{\alpha} = \left[\frac{A}{r^2} - \frac{4B}{r^4} - \frac{A}{2\left(1-\nu\right)r^2} \right]\epsilon^{*}_{\alpha\beta}X_{\beta}  + \left[\frac{8B}{r^6} + \frac{A}{2\left(1-\nu\right)r^4}\right]X_{\alpha}\epsilon^{*}_{\beta\gamma}X_{\beta} X_{\gamma} \nonumber \\
&  = \left[\frac{A}{r^2}\frac{1-2\nu}{2\left(1-\nu\right)} - \frac{4B}{r^4}\right]\epsilon^{*}_{\alpha\beta}X_{\beta}  + \left[\frac{8B}{r^6} + \frac{A}{2\left(1-\nu\right)r^4}\right]X_{\alpha}\epsilon^{*}_{\beta\gamma}X_{\beta} X_{\gamma}
\end{eqnarray}
At $r=a$ (the radius of the circular inclusion), the form of expression Eq.~(\ref{26}) must match the form of the constrained displacement field of the inclusion which from Eq.~(\ref{5a}) is $\frac{3-4\nu}{4\left(1-\nu\right)}\epsilon^{*}_{\alpha\beta}X_{\beta}$. Thus the co-efficient of the second term in expression (\ref{26}) must go to zero at the inclusion boundary, which gives us
\begin{eqnarray}\label{27}
&\frac{8B}{a^6} + \frac{A}{2\left(1-\nu\right)a^4} = 0 \nonumber \\
&\Rightarrow B = \frac{-a^2A}{16\left(1-\nu \right)}
\end{eqnarray}
Thus we have
\begin{eqnarray}\label{28}
& u^{c}_{\alpha} =\left[\frac{A}{r^2}\frac{1-2\nu}{2\left(1-\nu\right)} - \frac{4}{r^4}\frac{-a^2A}{16\left(1-\nu \right)}\right]\epsilon^{*}_{\alpha\beta}X_{\beta}  + \left[\frac{8}{r^6}\frac{-a^2A}{16\left(1-\nu \right)} + \frac{A}{2\left(1-\nu\right)r^4}\right]X_{\alpha}\epsilon^{*}_{\beta\gamma}X_{\beta} X_{\gamma} \nonumber \\
& = \frac{A}{4r^2\left(1-\nu\right)} \left[2\left(1-2\nu\right) + \frac{a^2}{r^2}\right]\epsilon^{*}_{\alpha\beta}X_{\beta} + \frac{A}{2r^4\left(1-\nu\right)}\left[1 - \frac{a^2}{r^2}\right] X_{\alpha}\epsilon^{*}_{\beta\gamma}X_{\beta} X_{\gamma} \nonumber \\
\end{eqnarray}
And the value of $u^{c}_{\alpha}$ at $r=a$ should match the value obtained from Eq. (\ref{5a}), implying
\begin{eqnarray}\label{29}
 \frac{3-4\nu}{4\left(1-\nu\right)} = \frac{A}{4\left(1-\nu\right)a^2}\left[2\left(1-2\nu\right) + 1\right]
 \Rightarrow \frac{3-4\nu}{4\left(1-\nu\right)}  = \frac{A\left(3-4\nu\right)}{4\left(1-\nu\right)a^2} \Rightarrow A = a^2
\end{eqnarray}
The expression for $u^{c}_{\alpha}$ becomes:
\begin{equation}\label{30}
u^{c}_{\alpha} = \frac{1}{4\left(1-\nu\right)}\left(\frac{a^2}{r^2}\right) \left[2\left(1-2\nu\right) + \left(\frac{a^2}{r^2}\right)\right]\epsilon^{*}_{\alpha\beta}X_{\beta} + \frac{1}{2\left(1-\nu\right)}\left(\frac{a^2}{r^2}\right) \left[1 - \left(\frac{a^2}{r^2}\right)\right] \epsilon^{*}_{\beta\gamma} \frac{X_{\alpha} X_{\beta} X_{\gamma}}{r^2}
\end{equation}
From Eq.~\ref{1}, we have $\epsilon _{\alpha\beta}^{*} = \epsilon^{*}\left(2\hat n_{\alpha}\hat n_{\beta} - \delta_{\alpha\beta}\right)$ and thus
\begin{eqnarray}\label{31}
& \epsilon _{\alpha\beta}^{*}X_{\beta} = \epsilon^{*}\left[2\hat n_{\alpha}\left(\mathbf{\hat n} \cdot \vec X\right) - X_{\alpha}\right] \nonumber \\
&X_{\beta}\epsilon _{\beta\gamma}^{*}X_{\gamma} = \epsilon^{*}X_{\beta}\left(2\hat n_{\beta}\hat n_{\gamma} - \delta_{\beta\gamma}\right)X_{\gamma} = \epsilon^{*}\left[2\left(\mathbf{\hat n} \cdot \vec X\right)^2 - r^2\right]
\end{eqnarray}
allowing us to write the final vectorial expression for the displacement field:
\begin{equation}\label{32}
  \addtolength{\fboxsep}{5pt}
   \begin{gathered}
\vec u^{c}\left(\vec X\right) = \frac{\epsilon^{*}}{4\left(1-\nu\right)}\left(\frac{a^2}{r^2}\right) \left[2\left(1-2\nu\right) + \left(\frac{a^2}{r^2}\right)\right]\left[2\mathbf{\hat n}\left(\mathbf{\hat n} \cdot \vec X\right) - \vec X\right] \\
 + \frac{\epsilon^{*}}{2\left(1-\nu\right)}\left(\frac{a^2}{r^2}\right) \left[1 - \left(\frac{a^2}{r^2}\right)\right] \left[\frac{2\left(\mathbf{\hat n} \cdot \vec X\right)^2}{r^2} - 1\right]\vec X
\end{gathered}
\end{equation}\label{32a}

We can also derive expressions for the constrained stress and strain fields. Noting that
\begin{eqnarray}
\frac{\partial f(r)}{\partial X_{\beta}} &=& f'(r) \frac{\partial r}{\partial X_{\beta}} = f'(r)\frac{X_{\beta}}{r} \nonumber \\
\frac{\partial\left(\mathbf{\hat n}\cdot \vec X\right)}{\partial X_{\beta}} &=& \hat n_{\beta}
\end{eqnarray}
we obtain
\begin{eqnarray}\label{32b}
\begin{gathered}
\frac{\partial u^{c}_{\alpha}}{\partial X_{\beta}} = \frac{\partial}{\partial X_{\beta}}\bigg[\frac{\epsilon^{*}}{4\left(1-\nu\right)}\left(\frac{a^2}{r^2}\right) \left\lbrace2\left(1-2\nu\right) + \left(\frac{a^2}{r^2}\right)\right\rbrace\left\lbrace2\hat n_{\alpha}\left(\mathbf{\hat n} \cdot \vec X\right) - X_{\alpha}\right\rbrace \\
 + \frac{\epsilon^{*}}{2\left(1-\nu\right)}\left(\frac{a^2}{r^2}\right) \left\lbrace 1 - \left(\frac{a^2}{r^2}\right)\right\rbrace \left\lbrace \frac{2\left(\mathbf{\hat n} \cdot \vec X\right)^2}{r^2} - 1\right\rbrace X_{\alpha}\bigg] \\\\
 = \frac{\epsilon^{*}}{4\left(1-\nu\right)}\left\lbrace-4\left(1-2\nu\right)\left(\frac{a^2}{r^4}\right) -4\left(\frac{a^4}{r^6}\right)\right\rbrace\left\lbrace2\hat n_{\alpha}\left(\mathbf{\hat n} \cdot \vec X\right) - X_{\alpha}\right\rbrace X_{\beta} \\
+ \frac{\epsilon^{*}}{4\left(1-\nu\right)} \left\lbrace2\left(1-2\nu\right)\left(\frac{a^2}{r^2}\right) + \left(\frac{a^4}{r^4}\right)\right\rbrace\left\lbrace2\hat n_{\alpha}\hat n_{\beta} - \delta_{\alpha\beta}\right\rbrace \\
 + \frac{\epsilon^{*}}{2\left(1-\nu\right)}\left\lbrace \left(-\frac{2a^2}{r^4}\right) + \left(\frac{4a^4}{r^6}\right)\right\rbrace \left\lbrace \frac{2\left(\mathbf{\hat n} \cdot \vec X\right)^2}{r^2} - 1\right\rbrace X_{\alpha}X_{\beta} \\
 + \frac{\epsilon^{*}}{2\left(1-\nu\right)}\left\lbrace \left(\frac{a^2}{r^2}\right) - \left(\frac{a^4}{r^4}\right)\right\rbrace \left\lbrace \frac{4\left(\mathbf{\hat n} \cdot \vec X\right)n_{\beta}}{r^2}  - \frac{4\left(\mathbf{\hat n} \cdot \vec X\right)^2X_{\beta}}{r^4} \right\rbrace X_{\alpha} \\
 + \frac{\epsilon^{*}}{2\left(1-\nu\right)}\left\lbrace \left(\frac{a^2}{r^2}\right) - \left(\frac{a^4}{r^4}\right)\right\rbrace \left\lbrace \frac{2\left(\mathbf{\hat n} \cdot \vec X\right)^2}{r^2} - 1\right\rbrace \delta_{\alpha\beta} \\\\
\Rightarrow \frac{\partial u^{c}_{\alpha}}{\partial X_{\beta}} = \frac{\epsilon^*}{4\left(1-\nu\right)}\bigg[-4\left(\frac{a^2}{r^2}\right) \left\lbrace\left(1-2\nu\right) + \left(\frac{a^2}{r^2}\right)\right\rbrace\left\lbrace2\hat n_{\alpha}\frac{\left(\mathbf{\hat n} \cdot \vec X\right)}{r} - \frac{X_{\alpha}}{r}\right\rbrace \frac{X_{\beta}}{r} \\
+ \left(\frac{a^2}{r^2}\right)\left\lbrace2\left(1-2\nu\right) + \left(\frac{a^2}{r^2}\right)\right\rbrace\left\lbrace2\hat n_{\alpha}\hat n_{\beta} - \delta_{\alpha\beta}\right\rbrace \\
 -4 \left(\frac{a^2}{r^2}\right)\left\lbrace 1 - 2\left(\frac{a^2}{r^2}\right)\right\rbrace \left\lbrace \frac{2\left(\mathbf{\hat n} \cdot \vec X\right)^2}{r^2} - 1\right\rbrace \frac{X_{\alpha}X_{\beta}}{r^2} \\
+ 8\left(\frac{a^2}{r^2}\right)\left\lbrace 1 - \left(\frac{a^2}{r^2}\right)\right\rbrace \left\lbrace \frac{\left(\mathbf{\hat n} \cdot \vec X\right)\hat n_{\beta}}{r}  - \frac{\left(\mathbf{\hat n} \cdot \vec X\right)^2}{r^2}\frac{X_{\beta}}{r} \right\rbrace \frac{X_{\alpha}}{r} \\
+ 2\left(\frac{a^2}{r^2}\right)\left\lbrace 1 - \left(\frac{a^2}{r^2}\right)\right\rbrace \left\lbrace \frac{2\left(\mathbf{\hat n} \cdot \vec X\right)^2}{r^2} - 1\right\rbrace \delta_{\alpha\beta}\bigg]
\end{gathered}
\end{eqnarray}
Thus the expression for $\epsilon^{c}_{\alpha\beta} \equiv \frac{1}{2}\left(\frac{\partial u^{c}_{\alpha}}{\partial X_{\beta}} + \frac{\partial u^{c}_{\beta}}{\partial X_{\alpha}}\right)$ is
\begin{eqnarray}\label{32c}
\begin{gathered}
\epsilon^{c}_{\alpha\beta}(\vec{X}) = \frac{\epsilon^*}{8\left(1-\nu\right)}\bigg[-4\left(\frac{a^2}{r^2}\right) \left\lbrace\left(1-2\nu\right) + \left(\frac{a^2}{r^2}\right)\right\rbrace\left\lbrace2\left(\frac{\mathbf{\hat n} \cdot \vec X}{r}\right)\left(\frac{\hat n_{\alpha} X_{\beta}}{r} + \frac{\hat n_{\beta} X_{\alpha}}{r}\right) - 2\frac{X_{\alpha} X_{\beta}}{r^2}\right\rbrace \\
+ 2\left(\frac{a^2}{r^2}\right)\left\lbrace2\left(1-2\nu\right) + \left(\frac{a^2}{r^2}\right)\right\rbrace\left\lbrace2\hat n_{\alpha}\hat n_{\beta} - \delta_{\alpha\beta}\right\rbrace \\
-8 \left(\frac{a^2}{r^2}\right)\left\lbrace 1 - 2\left(\frac{a^2}{r^2}\right)\right\rbrace \left\lbrace \frac{2\left(\mathbf{\hat n} \cdot \vec X\right)^2}{r^2} - 1\right\rbrace \frac{X_{\alpha}X_{\beta}}{r^2} \\
+ 8\left(\frac{a^2}{r^2}\right)\left\lbrace 1 - \left(\frac{a^2}{r^2}\right)\right\rbrace \left\lbrace \left(\frac{\mathbf{\hat n} \cdot \vec X}{r}\right)\left(\frac{X_{\alpha} \hat n_{\beta}}{r} + \frac{X_{\beta} \hat n_{\alpha}}{r}\right)  - 2\frac{\left(\mathbf{\hat n} \cdot \vec X\right)^2}{r^2}\frac{X_{\alpha} X_{\beta}}{r^2} \right\rbrace \\
+4\left(\frac{a^2}{r^2}\right)\left\lbrace 1 - \left(\frac{a^2}{r^2}\right)\right\rbrace \left\lbrace \frac{2\left(\mathbf{\hat n} \cdot \vec X\right)^2}{r^2} - 1\right\rbrace \delta_{\alpha\beta}\bigg]
\end{gathered}
\end{eqnarray}
allowing us to write the final expression for the constrained strain in the matrix
\begin{eqnarray}\label{32d}
\begin{gathered}
\epsilon^{c}_{\alpha\beta}(\vec{X}) = \frac{\epsilon^*}{4\left(1-\nu\right)}\bigg[-4\left(\frac{a^2}{r^2}\right) \left\lbrace\left(1-2\nu\right) + \left(\frac{a^2}{r^2}\right)\right\rbrace\left\lbrace\left(\frac{\mathbf{\hat n} \cdot \vec X}{r}\right)\left(\frac{\hat n_{\alpha} X_{\beta}}{r} + \frac{\hat n_{\beta} X_{\alpha}}{r}\right) - \frac{X_{\alpha} X_{\beta}}{r^2}\right\rbrace \\
+ \left(\frac{a^2}{r^2}\right)\left\lbrace2\left(1-2\nu\right) + \left(\frac{a^2}{r^2}\right)\right\rbrace\left\lbrace2\hat n_{\alpha}\hat n_{\beta} - \delta_{\alpha\beta}\right\rbrace \\
-4 \left(\frac{a^2}{r^2}\right)\left\lbrace 1 - 2\left(\frac{a^2}{r^2}\right)\right\rbrace \left\lbrace \frac{2\left(\mathbf{\hat n} \cdot \vec X\right)^2}{r^2} - 1\right\rbrace \frac{X_{\alpha}X_{\beta}}{r^2} \\
+ 4\left(\frac{a^2}{r^2}\right)\left\lbrace 1 - \left(\frac{a^2}{r^2}\right)\right\rbrace \left\lbrace \left(\frac{\mathbf{\hat n} \cdot \vec X}{r}\right)\left(\frac{X_{\alpha} \hat n_{\beta}}{r} + \frac{X_{\beta} \hat n_{\alpha}}{r}\right)  - 2\frac{\left(\mathbf{\hat n} \cdot \vec X\right)^2}{r^2}\frac{X_{\alpha} X_{\beta}}{r^2} \right\rbrace \\
+2\left(\frac{a^2}{r^2}\right)\left\lbrace 1 - \left(\frac{a^2}{r^2}\right)\right\rbrace \left\lbrace \frac{2\left(\mathbf{\hat n} \cdot \vec X\right)^2}{r^2} - 1\right\rbrace \delta_{\alpha\beta}\bigg]
\end{gathered}
\end{eqnarray}
The trace of $\epsilon^{c}_{\alpha\beta}$ is not zero in the elastic medium and  is given by
\begin{eqnarray}\label{32e}
\epsilon^{c}_{\eta\eta} &=& \frac{\epsilon^*}{4\left(1-\nu\right)}\bigg[-4\left(\frac{a^2}{r^2}\right) \left\lbrace\left(1-2\nu\right) + \left(\frac{a^2}{r^2}\right)\right\rbrace\left\lbrace2\left(\frac{\mathbf{\hat n} \cdot \vec X}{r}\right)^2 - 1\right\rbrace \nonumber \\
&+& 0 \nonumber \\
&-& 4 \left(\frac{a^2}{r^2}\right)\left\lbrace 1 - 2\left(\frac{a^2}{r^2}\right)\right\rbrace \left\lbrace \frac{2\left(\mathbf{\hat n} \cdot \vec X\right)^2}{r^2} - 1\right\rbrace \nonumber \\
&+& 0 \nonumber \\
&+&4\left(\frac{a^2}{r^2}\right)\left\lbrace 1 - \left(\frac{a^2}{r^2}\right)\right\rbrace \left\lbrace \frac{2\left(\mathbf{\hat n} \cdot \vec X\right)^2}{r^2} - 1\right\rbrace \bigg] \nonumber \\\\
\Rightarrow \epsilon^{c}_{\eta\eta} &=& \frac{\epsilon^*}{4\left(1-\nu\right)}\bigg[-4\left(\frac{a^2}{r^2}\right) \left\lbrace\left(1-2\nu\right) + \left(\frac{a^2}{r^2}\right)\right\rbrace\left\lbrace2\left(\frac{\mathbf{\hat n} \cdot \vec X}{r}\right)^2 - 1\right\rbrace \nonumber \\
&+&4\left(\frac{a^2}{r^2}\right)\left(\frac{a^2}{r^2}\right)\left\lbrace \frac{2\left(\mathbf{\hat n} \cdot \vec X\right)^2}{r^2} - 1\right\rbrace \bigg] \nonumber \\
\Rightarrow \epsilon^{c}_{\eta\eta} &=& -\epsilon^*\left(\frac{1-2\nu}{1-\nu}\right)\left(\frac{a^2}{r^2}\right)\left\lbrace \frac{2\left(\mathbf{\hat n} \cdot \vec X\right)^2}{r^2} - 1\right\rbrace
\end{eqnarray}

We are now in a position to calculate the constrained stress in the elastic medium due to the deformed inclusion.
It is given by the expression
\begin{eqnarray}\label{32f}
\sigma^{c}_{ij} = \frac{\C E}{1+\nu}\epsilon^{c}_{\alpha\beta} + \frac{\C E\nu}{\left(1+\nu\right)\left(1-2\nu\right)}\epsilon^{c}_{\eta\eta}\delta_{\alpha\beta} = \frac{\C E\epsilon^*}{4\left(1-\nu^2\right)}\bigg[.....\bigg]
- \frac{\C E\nu\epsilon^*}{\left(1-\nu^2\right)}\left(\frac{a^2}{r^2}\right)\left\lbrace \frac{2\left(\mathbf{\hat n} \cdot \vec X\right)^2}{r^2} - 1\right\rbrace\delta_{\alpha\beta}
\end{eqnarray}
where $\bigg[.....\bigg]$ is the expression inside the square brackets in Eq.~\ref{32d}.\\

\subsection{Constrained displacement field - Cartesian components}

\noindent It proves useful to have the explicit cartesian components of the displacement field for computational and graphical purposes. If we consider the unit-vector $\hat n$ making an angle of $\phi$ with the positive direction of the x-axis, then the cartesian components of equation \ref{32} are :
\begin{equation}\label{33}
  \addtolength{\fboxsep}{5pt}
   \begin{gathered}
u^{c}_x = \frac{\epsilon^{*}}{4\left(1-\nu\right)}\left(\frac{a^2}{r^2}\right) \left[2\left(1-2\nu\right) + \left(\frac{a^2}{r^2}\right)\right]\left[2\cos\phi\left(x\cos\phi + y\sin\phi\right) - x\right] \\
 + \frac{\epsilon^{*}}{2\left(1-\nu\right)}\left(\frac{a^2}{r^2}\right) \left[1 - \left(\frac{a^2}{r^2}\right)\right]\left[\frac{2\left(x\cos\phi + y\sin\phi\right)^2}{r^2} - 1\right]x \\
\Rightarrow u^{c}_x = \frac{\epsilon^{*}}{4\left(1-\nu\right)}\left(\frac{a^2}{r^2}\right) \left[2\left(1-2\nu\right) + \left(\frac{a^2}{r^2}\right)\right]\left[x\cos2\phi + y\sin2\phi\right] \\
+ \frac{\epsilon^{*}}{2\left(1-\nu\right)}\left(\frac{a^2}{r^2}\right) \left[1 - \left(\frac{a^2}{r^2}\right)\right]\left[\frac{\left(x^2-y^2\right)\cos2\phi + 2xy\sin2\phi}{r^2}\right]x \\\\
u^{c}_y = \frac{\epsilon^{*}}{4\left(1-\nu\right)}\left(\frac{a^2}{r^2}\right) \left[2\left(1-2\nu\right) + \left(\frac{a^2}{r^2}\right)\right]\left[2\sin\phi\left(x\cos\phi + y\sin\phi\right) - y\right] \\
 + \frac{\epsilon^{*}}{2\left(1-\nu\right)}\left(\frac{a^2}{r^2}\right) \left[1 - \left(\frac{a^2}{r^2}\right)\right]\left[\frac{2\left(x\cos\phi + y\sin\phi\right)^2}{r^2} - 1\right]y \\
\Rightarrow u^{c}_y = \frac{\epsilon^{*}}{4\left(1-\nu\right)}\left(\frac{a^2}{r^2}\right) \left[2\left(1-2\nu\right) + \left(\frac{a^2}{r^2}\right)\right]\left[x\sin2\phi - y\cos2\phi\right] \\
+ \frac{\epsilon^{*}}{2\left(1-\nu\right)}\left(\frac{a^2}{r^2}\right) \left[1 - \left(\frac{a^2}{r^2}\right)\right]\left[\frac{\left(x^2-y^2\right)\cos2\phi + 2xy\sin2\phi}{r^2}\right]y
\end{gathered}
\end{equation}

\section{Calculation of the energy of N quadrupoles}

Eq.~(\ref{etot}) can be re-written using $\epsilon_{\alpha\beta} \equiv 1/2\left(u_{\alpha,\beta} +  u_{\beta,\alpha}\right)$ as
\begin{eqnarray}
E = \frac{1}{4}\sum_{i=1}^{N}\int_{V_{0}^i}\sigma^{(i)}_{\alpha\beta}\left(u_{\alpha,\beta}^{(i)} +  u_{\beta,\alpha}^{(i)}\right)dV + \frac{1}{4}\int_{V -  \sum_{i=1}^{N}V_0^{(i)}}\sigma^{(m)}_{\alpha\beta}\left(u_{\alpha,\beta}^{(m)} +  u_{\beta,\alpha}^{(m)}\right)dV
\end{eqnarray}
Using the symmetry of the stress tensor, we obtain
\begin{equation}\label{Estart}
E = \frac{1}{2}\sum_{i=1}^{N}\int_{V_{0}^i}\sigma^{(i)}_{\alpha\beta}u^{(i)}_{\beta,\alpha}dV + \frac{1}{2}\int_{V -  \sum_{i=1}^{N}V_0^{(i)}}\sigma^{(m)}_{\alpha\beta}u_{\beta,\alpha}^{(m)}dV
\end{equation}
We also have the identity
\begin{eqnarray}\label{iden}
\sigma_{\alpha\beta} u_{\beta,\alpha} = \left(\sigma_{\alpha\beta}u_{\beta}\right)_{,\alpha} - \sigma_{\alpha\beta,\alpha}u_{\beta} = \left(\sigma_{\alpha\beta}u_{\beta}\right)_{,\alpha}
\end{eqnarray}
if there are no body forces. Thus we can write Eq.~\ref{Estart} as
\begin{eqnarray}\label{E2}
E = \frac{1}{2}\sum_{i=1}^{N}\int_{V_{0}^{(i)}}\left(\sigma^{(i)}_{\alpha\beta}u_{\beta}^{(i)}\right)_{,\alpha}dV + \frac{1}{2}\int_{V -  \sum_{i=1}^{N}V_0^{(i)}}\left(\sigma^{(m)}_{\alpha\beta}u_{\beta}^{(m)}\right)_{,\alpha}dV
\end{eqnarray}
Using Gauss's theorem to convert these volume integrals into area integrals, we obtain
\begin{eqnarray}\label{E3}
E = \frac{1}{2}\sum_{i=1}^{N}\int_{S_{0}^i}\sigma^{(i)}_{\alpha\beta}u_{\beta}^{(i)}\hat n^{(i)}_{\alpha} dS - \sum_{i=1}^N\frac{1}{2}\int_{S_0^i}\sigma^{(m)}_{\alpha\beta}u_{\beta}^{(m)}\hat n^{(i)}_{\alpha}dS + \frac{1}{2}\int_{S_{\infty}}\sigma^{(m)}_{\alpha\beta}u_{\beta}^{(m)}\hat n^{(\infty)}_{\alpha}dS
\end{eqnarray}
where $\mathbf{\hat n}^{(i)}$ and $\mathbf{\hat n}^{(\infty)}$ are unit normal vectors \textit{both} pointing outwards respectively from the inclusion volume $V_0^i$ and the matrix boundary. Eq.~\ref{E3} can be rewritten as follows
\begin{eqnarray}\label{E4}
\begin{gathered}
E = \frac{1}{2}\int_{S^{(\infty)}}\sigma^{(m)}_{\alpha\beta}u_{\beta}^{(m)}\hat n^{(\infty)}_{\alpha}dS   +\frac{1}{2}\sum_{i=1}^N\int_{S_0^{(i)}}\left(\sigma^{(i)}_{\alpha\beta}u_{\beta}^{(i)} - \sigma^{(m)}_{\alpha\beta}u_{\beta}^{(m)}\right)\hat n^{(i)}_{\alpha}dS \\
\Rightarrow E = \frac{1}{2}\sigma^{(\infty)}_{\alpha\beta}\epsilon_{\beta\gamma}^{(\infty)}\int_{S^{(\infty)}}X_{\gamma}\hat n^{(\infty)}_{\alpha}dS   +\frac{1}{2}\sum_{i=1}^N\int_{S_0^{(i)}}\left(\sigma^{(i)}_{\alpha\beta}u_{\beta}^{(i)} - \sigma^{(m)}_{\alpha\beta}u_{\beta}^{(m)}\right)\hat n^{(i)}_{\alpha}dS \\
\Rightarrow E = \frac{1}{2}\sigma^{\infty}_{\alpha\beta}\epsilon_{\beta\alpha}^{\infty}V +\frac{1}{2}\sum_{i=1}^N\int_{S_0^{(i)}}\left(\sigma^{(i)}_{\alpha\beta}u_{\beta}^{(i)} - \sigma^{(m)}_{\alpha\beta}u_{\beta}^{(m)}\right)\hat n^{(i)}_{\alpha}dS
\end{gathered}
\end{eqnarray}
\end{widetext}

Thus we can write using the expressions earlier written down in Eq.~(\ref{epsm})
\begin{eqnarray}\label{epsm2}
\begin{gathered}
\epsilon^{(m)}_{\alpha\beta}\left(\vec X\right) = \epsilon^{(\infty)}_{\alpha\beta} + \sum_{i=1}^N \epsilon_{\alpha\beta}^{(c,i)}\left(\vec X\right) \\
\sigma^{(m)}_{\alpha\beta}\left(\vec X\right) = \sigma^{(\infty)}_{\alpha\beta} + \sum_{i=1}^N \sigma_{\alpha\beta}^{(c,i)}\left(\vec X\right) \\
u^{(m)}_{\alpha}\left(\vec X\right) = u^{(\infty)}_{\alpha}(\vec{X}) + \sum_{i=1}^N u_{\alpha}^{(c,i)}\left(\vec X\right)
\end{gathered}
\end{eqnarray}
where $\epsilon_{\alpha\beta}^{(c,i)}\left(\vec X\right)$ indicates the constrained strain at location $\vec X$ in the matrix due to the eshelby labeled with the index $i$ etc. We also have for locations $\vec{X}$ inside the inclusions
\begin{eqnarray}\label{epsi}
\begin{gathered}
\epsilon^{(i)}_{\alpha\beta}\left(\vec X\right) = \epsilon^{(\infty)}_{\alpha\beta} + \sum_{j\neq i}\epsilon^{(c,j)}\left(\vec X\right) + \epsilon_{\alpha\beta}^{(c,i)} - \epsilon^{(*,i)}_{\alpha\beta}\\
\sigma^{(i)}_{\alpha\beta}\left(\vec X\right) = \sigma^{(\infty)}_{\alpha\beta} + \sum_{j\neq i}\sigma^{(c,j)}\left(\vec X\right) + \sigma_{\alpha\beta}^{(c,i)} - \sigma^{(*,i)} \\
u^{(i)}_{\alpha}\left(\vec X\right) = u^{(\infty)}_{\alpha} + \sum_{j\neq i}u^{(c,j)}_{\alpha}\left(\vec X\right) + u_{\alpha}^{(c,i)} - \epsilon^{(*,i)}_{\alpha\beta}X_{\beta}
\end{gathered}
\end{eqnarray}
where $\epsilon^{(*,i)}$ is the eigenstrain of the $i$th Eshelby and so on. Note that in the expression for the  strain in the inclusion given by Eq.~(\ref{epsi}) we have removed the eigenstrain from the constrained strain  $\epsilon_{\alpha\beta}^{(c,i)} - \epsilon^{(*,i)}_{\alpha\beta}$ leaving only the elastic contribution  in order to calculate correctly the elastic contribution to the energy. Using these expressions, the elastic energy of the system can be written from Eq. (\ref{E4})
\begin{eqnarray}\label{E5}
E& = &\frac{1}{2}\sigma^{(\infty)}_{\alpha\beta}\epsilon_{\beta\alpha}^{(\infty)}V \\&+&\frac{1}{2}\sum_{i=1}^N\int_{S_0^{(i)}}\left(\sigma^{(i)}_{\alpha\beta}u_{\beta}^{(i)} - \sigma^{(m)}_{\alpha\beta}u_{\beta}^{(m)}\right)\hat n^{(i)}_{\alpha}dS \ . \nonumber
\end{eqnarray}
Since the traction force has to be continuous at the inclusion boundary (Newton's third law), we have
\begin{eqnarray}\label{sigmai}
\sigma^{(i)}_{\alpha\beta}\hat n^{(i)}_{\alpha} = \sigma^{(m)}_{\alpha\beta}\hat n^{(i)}_{\alpha} \quad \text{at the inclusion boundary}
\end{eqnarray}
which gives us from Eq. (\ref{E5}),
\begin{eqnarray}\label{E6}
\!\!\!\!\!E \!\!= \!\!\frac{1}{2}\sigma^{(\infty)}_{\alpha\beta}\epsilon_{\beta\alpha}^{(\infty)}V \!+\!\frac{1}{2}\sum_{i=1}^N\int_{S_0^{(i)}}\!\sigma^{(i)}_{\alpha\beta}\hat n^{(i)}_{\alpha}\!\left(\!u_{\beta}^{(i)}\! -\! u_{\beta}^{(m)}\!\right)\!dS
\end{eqnarray}
We also have from Eqs. (\ref{epsm2}) and (\ref{epsi}),
\begin{eqnarray}\label{E7}
u^{(i)}_{\beta} - u^{(m)}_{\beta} = - \epsilon^{(*,i)}_{\beta\nu}X_{\nu} .
\end{eqnarray}
On plugging this expression into Eq.~(\ref{E6}) gives us finally
\begin{eqnarray}\label{46}
\begin{gathered}
E = \frac{1}{2}\sigma^{(\infty)}_{\alpha\beta}\epsilon_{\beta\alpha}^{(\infty)}V -\frac{1}{2}\sum_{i=1}^N\int_{S_0^{(i)}}\sigma^{(i)}_{\alpha\beta}\hat n^{(i)}_{\alpha}\epsilon^{(*,i)}_{\beta\nu}X_{\nu}dS \\
\Rightarrow  = \frac{1}{2}\sigma^{(\infty)}_{\alpha\beta}\epsilon_{\beta\alpha}^{(\infty)}V -\frac{1}{2}\sum_{i=1}^N\epsilon^{(*,i)}_{\beta\nu}\int_{V_0^{(i)}}\left(\sigma^{(i)}_{\alpha\beta}X_{\nu}\right)_{,\alpha}dV \\
\Rightarrow  = \frac{1}{2}\sigma^{(\infty)}_{\alpha\beta}\epsilon_{\beta\alpha}^{(\infty)}V -\frac{1}{2}\sum_{i=1}^N\epsilon^{(*,i)}_{\beta\nu}\int_{V_0^{(i)}}\sigma^{(i)}_{\alpha\beta}\delta_{\nu\alpha}dV \\
\Rightarrow  E = \frac{1}{2}\sigma^{(\infty)}_{\alpha\beta}\epsilon_{\beta\alpha}^{(\infty)}V -\frac{1}{2}\sum_{i=1}^NV_0^{(i)}\epsilon^{(*,i)}_{\beta\alpha}\overline{\sigma^{(i)}_{\alpha\beta}}
\end{gathered}
\end{eqnarray}
where $\overline{\sigma^{i}_{\alpha\beta}} \equiv (1/V_0^i)\int_{V_0^i}\sigma^{(i)}_{\alpha\beta}dV$. Using the expression for $\sigma^{(i)}_{\alpha\beta}$ from Eq.~(\ref{epsi}), we obtain
\begin{eqnarray}\label{47}
\overline{\sigma^{(i)}_{\alpha\beta}(\vec{\B X})} \approx \sigma^{(\infty)}_{\alpha\beta} + \sum_{j\neq i}\sigma^{(c,j)}\left(R_{ij}\right) + \sigma_{\alpha\beta}^{(c,i)} - \sigma^{(*,i)}.
\end{eqnarray}
Eq.~(\ref{47}) is a far field approximation that assumes that $R_{ij}\gg a$. As $R_{ij}\rightarrow a$ clearly the spatial integrals
contibuting to $\overline{\sigma^{c,i}_{\alpha\beta}}$ must be computed explicitly and cannot be replaced by the single distance $R_{ij}$ between the centers of the Eshelby inclusions $i$ and $j$.

\noindent Using expression~(\ref{46}) we obtain
\begin{widetext}
\begin{eqnarray}\label{48}
& E &=  \frac{1}{2}\sigma^{(\infty)}_{\alpha\beta}\epsilon_{\beta\alpha}^{(\infty)}V  - \frac{1}{2}\sigma^{(\infty)}_{\alpha\beta}\left(\sum_{i=1}^{N}\epsilon^{(*,i)}_{\beta\alpha}V_0^{(i)}\right) - \frac{1}{2}\sum_{i=1}^N\epsilon_{\beta\alpha}^{(*,i)}\sigma^{(c,i)}_{\alpha\beta}V_0^{(i)} + \frac{1}{2}\sum_{i=1}^N\epsilon^{(*,i)}_{\beta\alpha}\sigma_{\alpha\beta}^{(*,i)}V_0^{(i)} \nonumber \\
& &- \frac{1}{2} \sum_{i=1}^N \epsilon^{(*,i)}V_0^{(i)} \left(\sum_{j\neq i}\sigma_{\alpha\beta}^{(c,j)}\left(R_{ij}\right)\right) \nonumber \\
\Rightarrow E &=& E_{mat} + E^{\infty} + E_{esh} + E_{inc}.
\end{eqnarray}
where all these terms are defined in Eqs. (\ref{Emat})-(\ref{Einc}).
\end{widetext}

\section{The final form of $E_{\rm inc}$}
\begin{widetext}
 We can also explicitly write $E_{inc}$ showing its linear dependence on the eigenstrain by using Eq.~(\ref{Einc}). This gives us
\begin{eqnarray}\label{50}
E_{inc} = -\frac{\pi a^2}{2} \epsilon^*\sum_{\langle ij\rangle}\left[\left(2\hat n^{(i)}_{\alpha}\hat n^{(i)}_{\beta} - \delta_{\alpha\beta}\right)\sigma^{(c,j)}_{\alpha\beta}(R_{ij}) + \left(2\hat n^{(j)}_{\alpha}\hat n^{(j)}_{\beta} - \delta_{\alpha\beta}\right)\sigma^{(c,i)}_{\alpha\beta}(R_{ij})\right] .
\end{eqnarray}
Plugging Eq.~(\ref{32f}) into the above equation, we find that
the term inside the square braces in Eq.~(\ref{50}) can be written as:
\begin{eqnarray}\label{51}
&=& \frac{\C E\epsilon^*\left(2\hat n^{(j)}_{\alpha}\hat n^{(j)}_{\beta} - \delta_{\alpha\beta}\right)}{4\left(1-\nu^2\right)}\bigg[-4\left(\frac{a}{R_{ij}}\right)^2 \left\lbrace\left(1-2\nu\right) + \left(\frac{a}{R_{ij}}\right)^2\right\rbrace\nonumber \\
&\times&\left\lbrace\left(\frac{\mathbf{\hat n}^i \cdot \vec X^{ij}}{R_{ij}}\right)\left(\frac{\hat n^{(i)}_{\alpha} X^{(ij)}_{\beta}}{R_{ij}} + \frac{\hat n^{(i)}_{\beta} X^{(ij)}_{\alpha}}{R_{ij}}\right) - \frac{X^{(ij)}_{\alpha} X^{(ij)}_{\beta}}{(R_{ij})^2}\right\rbrace \nonumber \\
&+& \left(\frac{a}{R_{ij}}\right)^2\left\lbrace2\left(1-2\nu\right) + \left(\frac{a}{R_{ij}}\right)^2\right\rbrace\left\lbrace2\hat n^{(i)}_{\alpha}\hat n^{(i)}_{\beta} - \delta_{\alpha\beta}\right\rbrace \nonumber \\
&-&4 \left(\frac{a}{R_{ij}}\right)^2\left\lbrace 1 - 2\left(\frac{a}{R_{ij}}\right)^2\right\rbrace \left\lbrace \frac{2\left(\mathbf{\hat n}^{(i)} \cdot \vec X^{(ij)}\right)^2}{(R_{ij})^2} - 1\right\rbrace \frac{X^{(ij)}_{\alpha}X^{(ij)}_{\beta}}{(R_{ij})^2} \nonumber \\
&+& 4\left(\frac{a}{R_{ij}}\right)^2\left\lbrace 1 - \left(\frac{a}{R_{ij}}\right)^2 \right\rbrace \times \nonumber \\
&&\left\lbrace \left(\frac{\mathbf{\hat n}^{(i)} \cdot \vec X^{(ij)}}{R_{ij}}\right)\left(\frac{X^{(ij)}_{\alpha} \hat n^{(i)}_{\beta}}{R_{ij}} + \frac{X^{(ij)}_{\beta} \hat n^{(i)}_{\alpha}}{R_{ij}}\right) - 2\frac{\left(\mathbf{\hat n}^{(i)} \cdot \vec X^{(ij)}\right)^2}{(R_{ij})^2}\frac{X^{(ij)}_{\alpha} X^{(ij)}_{\beta}}{(R_{ij})^2} \right\rbrace \nonumber \\
&+&2\left(\frac{a}{R_{ij}}\right)^2\left\lbrace 1 - \left(\frac{a}{R_{ij}}\right)^2\right\rbrace \left\lbrace \frac{2\left(\mathbf{\hat n}^{(i)} \cdot \vec X^{(ij)}\right)^2}{(R_{ij})^2} - 1\right\rbrace \delta_{\alpha\beta}\bigg] \nonumber \\
&-& \frac{E\nu\epsilon^*}{1-\nu^2}\left(\frac{a^2}{r^2}\right)\left(\frac{2(\mathbf{\hat n}^{(i)} \cdot \vec X^{(ij)})^2}{r^2} - 1\right)\left(2\hat n^{(j)}_{\alpha}\hat n^{(j)}_{\beta} - \delta_{\alpha\beta}\right)\delta_{\alpha\beta} \nonumber \\
&+& \langle i \leftrightarrow j\rangle
\end{eqnarray}
where $\vec X^{(ij)}$ indicates the vector joining the centers of the eshelby pair labeled as $i$ and $j$, and $\langle i \leftrightarrow j\rangle$ in Eq.~(\ref{51}) represents the term obtained by exchanging $i$ and $j$.

\noindent In order to simplify Eq.~ (\ref{51}), we need the following identities
\begin{eqnarray}
&&\left(2\hat n^{(j)}_{\alpha}\hat n^{(j)}_{\beta} - \delta_{\alpha\beta}\right)\left\lbrace\left(\frac{\mathbf{\hat n}^{(i)} \cdot \vec X^{(ij)}}{R_{ij}}\right)\left(\frac{\hat n^{(i)}_{\alpha} X^{(ij)}_{\beta}}{R_{ij}} + \frac{\hat n^{(i)}_{\beta} X^{(ij)}_{\alpha}}{R_{ij}}\right) - \frac{X^{(ij)}_{\alpha} X^{(ij)}_{\beta}}{(R_{ij})^2}\right\rbrace + \langle i \leftrightarrow j\rangle \nonumber \\
&=& 8\hat n^{(i)}\cdot \hat n^{(j)} \left(\frac{\mathbf{\hat n}^{(i)}\cdot\vec X^{(ij)}}{R_{ij}}\right) \left(\frac{\mathbf{\hat n}^{(j)}\cdot\vec X^{(ij)}}{R_{ij}}\right) - 4\left(\frac{\mathbf{\hat n}^{(i)} \cdot \vec X^{(ij)}}{R_{ij}}\right)^2 - 4\left(\frac{\mathbf{\hat n}^{(j)} \cdot \vec X^{(ij)}}{R_{ij}}\right)^2 + 2 \\
&& \left(2\hat n^{(j)}_{\alpha}\hat n^{(j)}_{\beta} - \delta_{\alpha\beta}\right)\left(2\hat n^{(i)}_{\alpha}\hat n^{(i)}_{\beta} - \delta_{\alpha\beta}\right) + \langle i \leftrightarrow j\rangle = 4\left[2\left(\mathbf{\hat n}^{(i)}\cdot \mathbf{\hat n}^{(j)}\right)^2 - 1\right] \\
&& \left(2\hat n^{(j)}_{\alpha}\hat n^{(j)}_{\beta} - \delta_{\alpha\beta}\right)\frac{X^{(ij)}_{\alpha}X^{(ij)}_{\beta}}{(R_{ij})^2} + \langle i \leftrightarrow j\rangle = 2\left[2\left(\frac{\mathbf{\hat n}^{(j)} \cdot \vec X^{(ij)}}{R_{ij}}\right)^2 - 1\right] \\
&& \left(2\hat n^{(j)}_{\alpha}\hat n^{(j)}_{\beta} - \delta_{\alpha\beta}\right)\left\lbrace \left(\frac{\mathbf{\hat n}^{(i)} \cdot \vec X^{(ij)}}{R_{ij}}\right)\left(\frac{X^{(ij)}_{\alpha} \hat n^{(i)}_{\beta}}{R_{ij}} + \frac{X^{(ij)}_{\beta} \hat n^{(i)}_{\alpha}}{R_{ij}}\right)  - 2\frac{\left(\mathbf{\hat n}^{(i)} \cdot \vec X^{(ij)}\right)^2}{(R_{ij})^2}\frac{X^{(ij)}_{\alpha} X^{(ij)}_{\beta}}{(R_{ij})^2} \right\rbrace  \nonumber \\
&+& \langle i \leftrightarrow j\rangle \nonumber \\
&=& 8\left(\frac{\mathbf{\hat n}^{(i)}\cdot \vec X^{(ij)}}{R_{ij}}\right)\left(\frac{\mathbf{\hat n}^{(j)}\cdot \vec X^{(ij)}}{R_{ij}}\right)\left(\hat n^{(i)} \cdot \hat n^{(j)}\right) - 8\left(\frac{\mathbf{\hat n}^{(i)}\cdot \vec X^{(ij)}}{R_{ij}}\right)^2\left(\frac{\mathbf{\hat n}^{(j)}\cdot \vec X^{(ij)}}{R_{ij}}\right)^2  \\
&& \left(2\hat n^{(j)}_{\alpha}\hat n^{(j)}_{\beta} - \delta_{\alpha\beta}\right)\delta_{\alpha\beta} + \langle i \leftrightarrow j\rangle = 0
\end{eqnarray}

 Using these identities, we can write the final expression for the interaction energy in the form
 shown in Eq. (\ref{Esubinc}).

 \end{widetext}

\end{document}